\documentclass[12pt,preprint]{aastex}

\begin{document}
\title {Infrared Properties of Close Pairs of Galaxies}
\author{Margaret J. Geller}
\affil{Smithsonian Astrophysical Observatory, 
\\ 60 Garden St., Cambridge, MA 02138}
\email{mgeller@cfa.harvard.edu}

\author{Scott J. Kenyon}
\affil{Smithsonian Astrophysical Observatory, 
\\ 60 Garden St., Cambridge, MA 02138}
\email{skenyon@cfa.harvard.edu}

\author{Elizabeth J. Barton}
\affil{Department of Physics and Astronomy, University of California Irvine, 
\\ 4154 Frederick Reines Hall, Irvine, CA 92697}
\email{ebarton@uci.edu}

\author{Thomas H. Jarrett}
\affil{Infrared Processing and Analysis Center,
\\ Spitzer Science Center, Jet Propulsion Laboratory,
\\ California Institute of Technology, Pasadena, CA 91125}
\email{jarrett@ipac.caltech.edu}

\author{Lisa J. Kewley}
\affil{Institute for Astronomy, University of Hawaii,
\\ 2680 Woodlawn Drive, Manoa, HI 96822} 
\email{kewley@ifa.hawaii.edu}

\begin{abstract}

We discuss spectroscopy and infrared photometry for a complete sample of $\sim$800
galaxies in close pairs objectively selected from the CfA2 redshift survey.
We use 2MASS to compare near infrared color-color diagrams for our sample with the Nearby
Field Galaxy Sample and with a set of IRAS flux-limited pairs 
from Surace et al. We construct a basic statistical model to explore the
physical sources of the substantial differences among these samples. The
model explains the spread of near infrared colors and is consistent with 
a picture where central star formation is triggered by the galaxy-galaxy
interaction before a merger occurs. For 160 galaxies we report new, deep JHK
photometry within our spectroscopic aperture and we use the combined
spectroscopic and photometric data to explore the physical conditions in the 
central bursts. We find a set of objects with H-K $\geq$ 0.45 and with a
large F$_{FIR}$/F$_H$. We interpret the very red H-K colors as evidence for
600---1000K dust within compact star-forming regions, perhaps similar to
super-star clusters identified in individual well-studied interacting
galaxies. The galaxies in our sample are
candidate ``hidden'' bursts or, possibly, ``hidden'' AGN. Over the entire pair sample,
both spectroscopic and photometric data show that the specific star formation
rate   decreases with the projected separation of the pair. The data suggest
that the near infrared color-color diagram is also a function of the
projected separation; all of the objects with central
near infrared colors indicative of bursts of star formation lie at small
projected separation. 

\end{abstract}


\section{INTRODUCTION}

Larson \& Tinsley (1978) first recognized the close connection between
galaxy-galaxy interactions and star formation. They
suggested that this
relationship is a fundamental ingredient of
galaxy formation and evolution. Studies of
individual
interacting systems and of statistical samples of close pairs in
the local universe have confirmed
and refined the initial results
(e.g. Kennicutt \& Keel 1984;
Kennicutt et al. 1987; Jones \& Stein 1989; Sekiguchi \& Wolstencroft 1992;
Keel 1993; Liu \& Kennicutt 1995a,b; Keel 1996; Donzelli \& Pastoriza 1997;
Barton, Geller, \& Kenyon 2000 (BGK hereafter); Barton Gillespie, Geller, \&
Kenyon (2003; BGK2 hereafter); Lambas et al. 2003; Allam et al. 2004;
Alonso et al. 2004; 
Nikolic et al. 2004). Examinations of the
universe at high redshift underscore the
importance of interactions in molding the evolution of galaxies (e.g.
Patton et al. 2002; Conselice et al. 2004; Lin et al. 2004; Papovich et al. 2005; Bell
et al. 2006).

The connection between interacting pairs of galaxies and
star formation is now well established observationally and there is a
promising correspondence between the predictions of simulations and
the properties of well-defined samples of close pairs. The data are
consistent with a burst of star formation triggered by a close
galaxy-galaxy interaction; the burst continues and ages as the galaxies move
apart (Mihos et al. 1991; Mihos \& Hernquist 1996; BGK, BGK2,
Lambas et al. 2003; Nikolic et al. 2004). The
primary feature of the data is an anti-correlation between the projected
separation of a pair on the sky and measures of the normalized star
formation rate. This anti-correlation suggests 
a dependence of the burst strength and probably burst age on
the separation of the galaxy pair (BGK2).

Here we focus on  near infrared observations of close pairs of
galaxies in the full BGK sample with attention to the distinctive features of 
infrared emission possibly associated with the interaction.
The literature contains detailed near infrared observations of
famous individual interacting galaxy pairs (e.g. Lopez-Sanchez et al. 2004)
and a few optical-infrared imaging surveys of small objectively selected
samples of close pairs.
Bushouse \& Werner (1990) imaged a sample of 22
interacting galaxies. In their sample, the nuclei are redder
than the outer regions by $\sim 0.4$
and 0.9 mag in J-K and R-K, respectively. They attribute the color gradient
mostly to an increase in the amount of dust in the nuclear regions and argue
that the infrared colors are unaffected by dust emission.
Cutri and McAlary (1985) carried out small aperture photometry of a larger
sample of pairs. They concluded that a larger fraction of interacting than
non-interacting galaxies have J-K and H-K colors outside the normal
range. Giuricin et al. (1993) used a compendium of data from the literature
to investigate the effect of interactions on the near infrared properties
of spiral galaxies. In contrast with
Bushouse and Werner (1990), they concluded that apparently
interacting galaxies have
H-K and K-L excesses indicative of thermal emission from
hot dust probably related to star formation induced by the interaction.
We use our spectroscopy and matched small aperture photometry
along with 2MASS photometry to 
revisit these issues in a much larger dataset of nearly 800 galaxies in
close pairs and n-tuples.

In the far infrared, IRAS observations still yield the largest datasets for
consideration of the properties of galaxy-galaxy interactions as a
function of observable descriptors of the pair (see, for example, Goto
2005). Telesco, Wolstencroft
\& Done (1988) compiled IRAS observations of pairs selected from
the catalog of Arp \& Madore (1987) and resolved by IRAS. In their set of
93 pairs of comparable
luminosity, they
find that pairs with the highest far infrared color temperature have the smallest
projected separation on the sky. They concluded that interactions
measurably enhance the intensity or efficiency of star formation.
Bushouse et al. (1988) reached similar conclusions based on IRAS
observations of a sample of pairs selected on the basis of optical
morphology.

Recently Surace, Sanders \& Mazzarella (2004; SSM hereafter) used HIRES image
reconstruction to resolve the IRAS emission at 12, 25, 60, and 100 $\mu$m
for 106 interacting galaxy systems where the galaxies are separated 
by less
than three average galaxy diameters. The sample has a flux limit at 60 $\mu$m
of 5.24 Jy. We use their sample to provide a context
for discussion of the  
generally unresolved IRAS detections in our sample. Neither the SSM or BGK
samples contain any ultraluminous infrared galaxies.

The complexity of the underlying sources of emission in the near infrared 
has probably been a disincentive to investigating pairs
in this spectral range. 
With current data it remains difficult to disentangle the
effects of reddening and thermal
dust emission at several temperatures. Photometry at L
together with spatially resolved photometry of
large samples at longer wavelengths
are necessary for clearer understanding of these issues. 
Currently available data are, however, adequate to demonstrate the
promise of the infrared for probing the galaxy-galaxy merger process
and the links between galaxy-galaxy interactions and star formation.

Here we compile the largest sample to date of near infrared colors of
galaxy pairs. The data we consider include 2MASS JHK$_s$
photometry for 791 galaxies in the
BGK sample and our own deeper JHK small aperture photometry for
a subsample of 160 BGK galaxies.
We also use  IRAS detections and upper limits 
for the entire BGK 
sample as a marker of the relationship between galaxy-galaxy interactions,
star formation, and thermal emission from hot dust
in these systems. We use our optical spectra and infrared colors measured 
through the same aperture to investigate basic properties of the central
regions of these galaxies.
 
Sec. 2 is a description of the infrared photometry and a review of the
optical photometry and spectroscopy. The data include
resolved 2MASS photometry for 44 systems unresolved in the public catalogs
and for 24 galaxies not detected or confused with nearby stars or galaxies.
In Section 3 we investigate the
infrared properties of the sample. We discuss
the near infrared color-color
diagram and investigate the relationships among
the near and far infrared colors, the Balmer decrement,
and the normalized star
formation rate. In Section 4 we show that the 
star formation rate as indicated by the normalized far infrared
flux increases as the projected pair separation decreases 
as expected if the interaction triggers star formation. There is
also a suggestive increase in the  spread of
near infrared color
as the projected separation of the pair decreases. This
increased  spread is evidence for triggered central bursts which may be either 
very blue or very  heavily obscured and reddened.  We
conclude in section 5.

\section{OBSERVATIONS AND DATA REDUCTION}

\subsection {The Sample of Pairs}

BGK identified 358 close pairs and n-tuples in the original
magnitude limited CfA2 redshift survey 
with m$_{Zw} \leq$ 15.5. The original CfA2North covers the declination range
$8.5^\circ < \delta_{B1950} <   44.5^\circ$ and
right ascension range 8h $ < \alpha_{B1950} < $ 17h and includes 6500 galaxies
(Geller \& Huchra 1989; Huchra et al. 1990; Huchra, Geller \& Corwin 1995).
The original CfA2South covers the region $-2.5^\circ < \delta_{B1950}
< 48^\circ$ and 20h $ < \alpha_{B1950} < $ 4h and
includes 4283 galaxies (Giovanelli \& Haynes 1985, 1989, 1993;
Giovanelli et al. 1986; Haynes et al. 1988; Wegner,
Haynes \& Giovanelli 1993; Huchra et al. 1999; see also Falco et al. 1999
for updated data for CfA2 North and South).

The pairs were
originally selected with line-of-sight velocity separations
$\Delta{\rm V} < 1000$ km s$^{-1}$, projected separations 
$\Delta{\rm D} <  50 h^{-1}$ kpc, and $cz > 2300$ km s$^{-1}$.
In this pair sample,  n-tuples result from pairs linked by common members as
``friends-of-friends.''
The 2300 km s$^{-1}$ limit excludes the Virgo cluster and limits the angular
sizes of the galaxies relative to the slit aperture we use for spectroscopy 
(Section 2.5). We use a Hubble constant 100 $h$ km s$^{-1}$ Mpc$^{-1}$
unless otherwise specified.

Ninety percent of the pairs have $\Delta{\rm V} < 500$ km s$^{-1}$,
comparable with the typical pairwise velocity dispersion in the 
redshift survey (Marzke et al. 1995). We repeated all of the analysis 
below for pair samples restricted to $\Delta{\rm V} < 500$ km s$^{-1}$
(89\% of the original sample) and $\Delta{\rm V} < 400$ km s$^{-1}$
(84\% of the original sample); there are no significant differences
in any results for any sample.

The pairs with  $\Delta{\rm V} > 500$ km s$^{-1}$
appropriately lie within
rich clusters where the local velocity
dispersion exceeds the mean and the local density contrast substantially
exceeds the mean;
we include these pairs for completeness and for consistency with our previous
analyses of this catalog. We use the BGK technique
based on an estimate of the galaxy overdensity in a 5h$^{-1}$ Mpc
sphere around each pair in the
sample to compute
the interloper fraction for the entire pairs sample, $\lesssim 20$\%.
By restricting the pairs sample to regions of low density contrast, we
also repeated our entire analysis for 
samples where the estimated interloper fraction is $\lesssim 10$\%. These 
restricted pair samples yield the same results we obtain for the full sample
and provide reassurance that interlopers are not responsible for salient
differences between the near infrared characteristics
of pairs and the general galaxy population.

The pairs in the full sample we use were selected without explicit bias in
morphology or environment. Updated coordinates and redshifts
modify the sample slightly
from the original one discussed in BGK; the sample we analyze here
contains 791 rather than the 786 galaxies in BGK. These galaxies lie in 306
pairs, 37 triples, 8 quadruples, and 7 quintuples. From here on, for
simplicity, we refer to all of the systems as pairs.

\subsection {2MASS photometry}

2MASS is an all-sky survey with uniform,
complete photometry (Nikolaev et al. 2000; Skrutskie et al. 2006) in three infrared bands
J, H, and K$_{\rm s}$. $K_{\rm s}$ is a modified version of
the K filter designed specifically to reduce thermal background. For most of our analysis, we use the 20 mag arcsec$^{-2}$ isophotal elliptical
aperture photometry from the final extended
source catalog (XSC; Jarrett et al. 2000). We compare
slit J,H,K magnitudes (Section 2.3) with the 2MASS 7$^{\prime\prime}$
aperture magnitudes.

We downloaded JHK$_{\rm s}$ from the 2MASS 
extended source catalog using the IRSA web interface at 
IPAC\footnote{http:$//$irsa.ipac.caltech.edu$/$}. All 
791 galaxies in the complete pairs sample are detected. 

Among the BGK pairs, 44 systems (12\%) are unresolved in 2MASS. The median
recessional velocity, $cz$, of
the BGK sample is only 5852 km s$^{-1}$. The unresolved pairs are a potentially serious
problem for trying to use the 2MASS catalog for pair selection even at this
depth and certainly as a basis for a deeper
sample. One of us (T.H.J.) reanalyzed the 2MASS
data to obtain magnitudes for the individual galaxies in the unresolved BGK
pairs. For each galaxy, Table 1 lists the J2000 coordinate (column 1),
the heliocentric radial velocity, $cz$ (column 2), the J, H, and K$_s$
7$^{\prime\prime}$ aperture magnitudes (columns 3, 4, 5 respectively) and
the J, H, and K$_s$ 20 mag arcsec$^{-2}$ isophotal elliptical magnitudes
(columns 6, 7, 8, respectively). We indicate the previously unresolved
objects with an asterisk in column 9. Twenty-four pair galaxies are either
undetected or confused with nearby stars or nearby non-pair galaxies.
T.H.J. also reanalyzed these pairs which are indicated by a + in column 9.   

For 2MASS galaxies with J $\approx$ 11--13, the photometric errors
are 
$\sigma_J \approx$ 0.02--0.04, $\sigma_H \approx$ 0.03--0.05, and
$\sigma_{K_s} \approx$ 0.04--0.06 for the isophotal magnitudes, and
$\sigma_J \approx$ 0.01--0.03, $\sigma_H \approx$ 0.02--0.04, and
$\sigma_{K_s} \approx$ 0.03--0.05 for the 7$^{\prime\prime}$ aperture 
magnitudes. To estimate
the uncertainty in the colors, we compute the average color differences
\begin{equation}
a_{\rm J-H} = <| (J-H)_7 - (J-H)_{iso} |>
\end{equation}
\noindent
and
\begin{equation}
a_{\rm H-K} = <| (H-K)_7 - (H-K)_{iso} |>
\end{equation}
\noindent
The subscript {\it 7} refers to the 7$^{\prime\prime}$ aperture colors
and the subscript {\it iso} refers to the isophotal colors.  
For galaxies with J $\approx$ 11--13, 
$a_{\rm J-H}$ = 0.026 and $a_{\rm H-K}$ = 0.034.
Thus, the uncertainty in the typical 2MASS color 
is $\sim$ 0.03 mag. This approach to the calculation of the error in the
colors follows a procedure recommended by Press et al. (1992).

To compare the near infrared properties of the BGK sample with the general
galaxy population, we use the Nearby Field Galaxy Sample
(NFGS: Jansen et al. 2000a, 2000b). The NFGS was also selected from the
Zwicky catalog, but the selection reproduces the galaxy luminosity function.
Because it is drawn from a magnitude limited survey, the BGK sample is
biased against the lowest luminosity galaxies sampled by the NFGS.
We extracted 2MASS magnitudes for the NFGS galaxies
and we compare the near infrared colors
of the BGK pairs with the appropriate sample of representative NFGS
galaxies. We restrict this comparison to the H-band luminosity range 
log(L$_H$) = 8.75 - 11.7 L$_\odot$ (H$_o$ = 73 km s$^{-1}$ Mpc$^{-1}$) covered by both
samples. There are 181 galaxies in the NFGS comparison sample.

\subsection { Near Infrared Photometry}

We acquired JHK images of 160 pair galaxies
in the sample of BGK2 along with standard
stars. These pairs are a random selection of the BGK pairs.
We used several NIR cameras at the Fred L. Whipple Observatory
1.2-m and the Kitt Peak National Observatory (KPNO) 2.1-m telescopes.
Table 2 summarizes pertinent details of the observing runs
including the dates (column 1), the telescope (column 2), the imager (column 3)
and the field of view (column 4).  Each
galaxy observation consists of five to nine 60 sec exposures,
dithered by 30--75\arcsec~in RA and Dec to allow for accurate
sky-subtraction and the elimination of bad pixels during data
reduction.  All-sky observations of 10--14 Elias et al. (1988) and UKIRT photometric
standards each clear night yielded accurate calibration constants and
extinction corrections.

Our approach to reducing the IR camera data is based on
previous experience with crowded Galactic fields with
extended emission from reflection nebulae and H~II regions
\citep[e.g.][]{Bar97,Whi97,Gom01,Bra02,Bal04}. Using a 
pipeline developed by W. Wyatt, we calibrated each frame with 
standard routines in NOAO IRAF\footnote{IRAF is distributed by 
the National Optical Astronomy Observatories, which is operated 
by the Association of Universities for Research in Astronomy, 
Inc., under contract to the National Science Foundation.}.
To generate flat-fields for each night, we median-filter all 
program frames using IMCOM, remove hot and dead pixels with a 
bad pixel mask, and normalize the median of the flat-field to 
unity.  After dividing linearized program frames by the 
flat-field, we sort images by their median sky levels and 
select 11--15 images with sky levels closest to the median sky 
level of each program field.  Median-filtered images of each 
set of 11--15 images, scaled by their median levels, yield 
sky frames for each program field.  Sky-subtracted images with 
bad pixels removed have a median background level of zero and 
noise levels of 20.0--21.0 mag arcsec$^{-2}$ at H.

The technique to derive sky frames has several advantages over
traditional methods. In ideal conditions where the sky background
varies slowly and monotonically through the night, our approach 
yields sky frames composed of images acquired close in time to the
program frame, as in standard reduction packages. When the sky 
background fluctuates erratically, our sky frames have fewer low
level background features than traditional sky frames and do not
require a DC offset to match background levels. Comparisons with
traditional sky flats show that our approach reduces photometric 
errors by 0.01--0.02 mag.

To construct combined images for each pair, we use DRIZZLE
\citep{fru02} in the STSDAS package within IRAF.  Our procedure
uses IMEXAM to derive (x,y) centers for each galaxy on each
frame, DRIZZLE to shift the images to a common center, and IMCOM
to construct a median-filter image of each galaxy from the set
of DRIZZLEd images. 

We derive broadband magnitudes using a simple FORTRAN program to 
sum the flux in a rectangular aperture which replicates the size 
and orientation of the slit used for spectroscopic observations 
(Section 2.5).  The program uses the (x,y) coordinates of the 
peak intensity to center the slit on each galaxy.  Tests indicate
uncertainties of 0.01--0.02 mag for $\pm$1 pixel (0.3--1.2 arcsec)
uncertainties in the slit position and $\pm$1 pixel uncertainties
in the size of the slit. 

We estimate photometric uncertainties for the survey from
repeat measurements.  The uncertainty in the photometric
calibration for each night is 0.01--0.03 mag. Multiple 
measurements of each galaxy within an observing run yield 
a typical uncertainty of $\pm$0.03 mag for galaxies with 
H = 13--15.  Repeat measurements acquired on different 
observing runs indicate a similar uncertainty.  The median 
offset between different observing runs is $\pm$0.02 mag.  
We thus conclude that the typical photometric uncertainty in our
slit magnitudes is 
$\sim$ 0.04 mag.

To estimate the uncertainty in the colors from our slit magnitudes, 
we derive colors assuming a fixed slit position for all three bands. 
Repeat measurements of galaxies throughout a single observing run and 
between observing runs yield identical color uncertainties of 
$\pm$0.04 mag in the J-H and H-K colors. This error estimate is smaller 
than the $\pm$0.06 mag uncertainty expected from simply adding the errors 
of individual bands in quadrature. However, we determine the
color at a fixed slit position 
for all three bands. Thus the error in the slit position
should not be included in the error in color.
We therefore reduce the error in the color
by removing the uncertainty in slit position relative to the 
spectroscopic aperture, $\sqrt{2(0.02)^2}$ $\approx$ 0.03 mag, and by the 
median offset in the photometric calibration between observing runs, 
$\sim$ 0.02 mag, which yields an expected error, $\approx$ 0.04 mag,
that agrees with the measured uncertainty. 

Table 3 lists the measurements. Column (1) is the J2000 galaxy coordinate,
column (2) is the slit length, L, in
arcseconds, column (3) is the slit width, w, in arcseconds, column
(4) is the slit position angle in degrees, and columns (4)-(6) are
the corresponding J, H, and K slit magnitudes respectively.
For nine galaxies without optical spectra (marked with an asterisk in
column 7 of Table 3), 
we adopted the median
slit length, 20$^{\prime\prime}$, and the typical position angle of 90$^\circ$. 
The colors of these galaxies span the range of J-H and H-K colors
of other galaxies in the sample.

\begin{figure}
\plotone{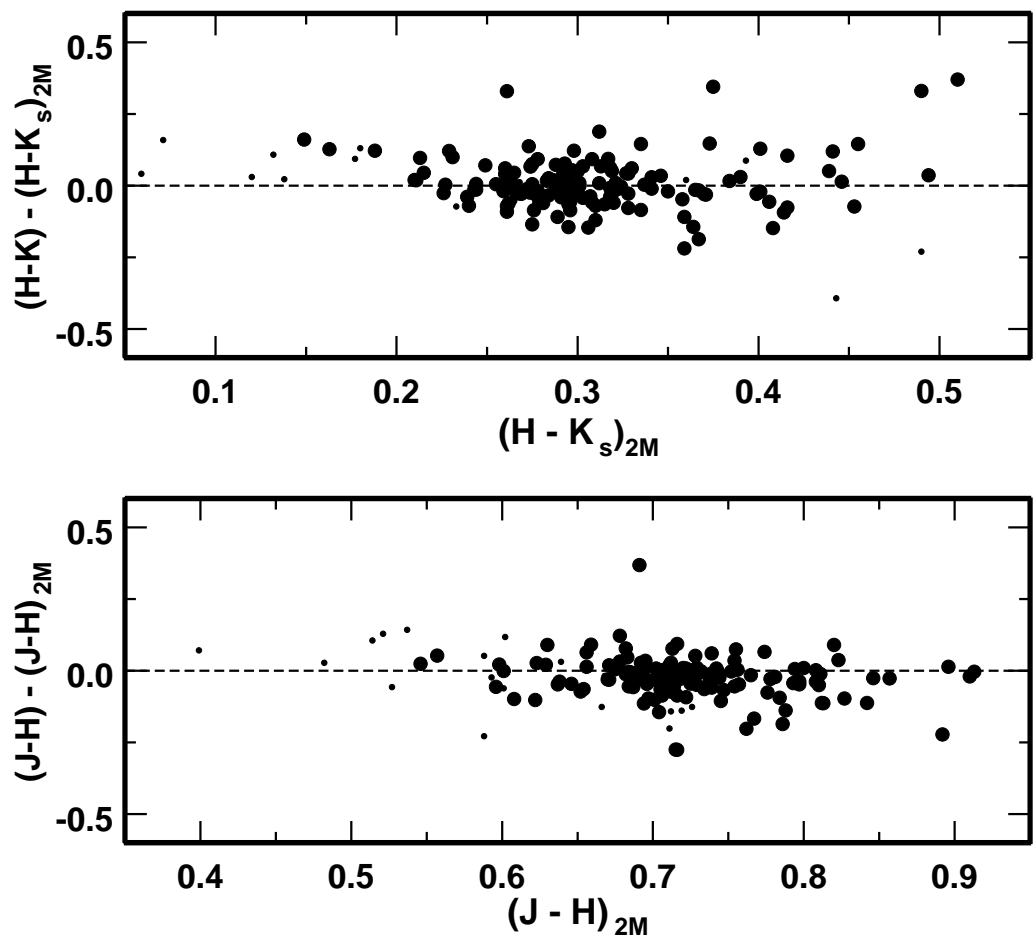}
\caption{Comparison of our slit aperture (J-H) and (H-K) photometry
with 2MASS
7$^{\prime\prime}$ diameter aperture (H-K$_s$)$_{2M}$ (top) and
(J-H)$_{2M}$ (bottom). The
small points are galaxies with 2MASS isophotal J $> 13$. The error in our colors
is 0.04 mag; the error in the 2MASS colors is 0.03 mag. The dispersion in the
relative colors is consistent with these errors. }
\label{fig:H-Kcomp}
\end{figure}

Fig. \ref{fig:H-Kcomp} compares the colors derived from our slit magnitude with 
2MASS colors derived from the 2MASS 7$^{\prime\prime}$ diameter magnitudes.
We use the 2MASS small aperture magnitudes here because they are the
best available match to our slit magnitudes.
The smaller points indicate galaxies with 2MASS  isophotal J$> 13$.
The mean color difference
$\langle{\rm (J - H) - (J - H)_{2M}}\rangle = -0.03\pm 0.005$;  
$\langle{\rm (H - K) - (H - K_s)_{2M}}\rangle = 0.01\pm 0.005$.
Removing the large outliers in Fig. \ref{fig:H-Kcomp}, the dispersion in
$\sigma[{\rm (J - H ) - (J - H)_{2M}}] = 0.05$ mag;
and $\sigma[{\rm (H - K)  - (H - K_s)_{2M}}] = 0.05$ mag. The
typical errors in the
2MASS colors are  0.03 mag: 
$\sigma[{\rm (J - H ) - (J - H)_{2M}}]$ and
$\sigma[{\rm (H - K)  - (H - K_s)_{2M}}]$ are thus consistent
with the 0.04 mag error we estimate in our slit aperture photometry.
These measures refer to the ensemble of points in the plots.
We suspect that the small blueward offset of our J-H relative to 2MASS 
results from differences in the aperture; our extracted apertures tend 
to be somewhat larger than 2MASS 7$^{\prime\prime}$  aperture. The reddest 
outliers in ${\rm (H - K) - (H - K_s)_{2M}}$ occur because our broader K 
filter provides greater sensitivity to emission from hot dust\footnote{With
a full-width at half maximum of 0.4 $\mu$m (2--2.4 $\mu$m), the standard 
K filter is 25\% wider than the K$_{\rm s}$ filter, which has a full-width 
at half maximum of 0.3 $\mu$m (2--2.3 $\mu$m). We acquired our K data in dry 
conditions, where K is more sensitive than $\rm K_s$ to hot dust emission with 
T = 600--1000 K.} (see Hunt et al.  2002 and Sections 3.1 and 3.2 below). 

\subsection {IRAS data}

We downloaded 12--100 $\mu$m photometry from the IRAS
extended and point source catalogs using the IRSA web interface
(Beichman et al. 1985).
These data yield 246 matches to pair galaxies, with 156 systems 
having both reliable 2MASS and 60 $\mu$m photometry. 
All but 10 of
these 156 systems have reliable 100 $\mu$m fluxes. 
In 5 pairs, IRAS unambiguously detected both galaxies; in 47 pairs, IRAS
detected one of the galaxies unambiguously; in 107 pairs, the IRAS detection
is unresolved.

Table 4 lists sample IRAS data for ten BGK pairs of galaxies.  We include 
a J2000 BGK coordinate designation (column 1), 
IRAS B1950 
coordinate designation (column 2), and the IRAS photometry (columns 3-6).  
The typical error in the 60 $\mu$m detections is 0.25 Jy; in the 100 $\mu$m
detections it is 0.5 Jy. The typical error in the upper limits at 60 $\mu$m
is 0.5 Jy.
In this Table, we associate the IRAS flux with the pair galaxy closest 
to the IRAS coordinate. In our discussion we treat the IRAS emission
as a property of system; the poor spatial resolution of IRAS prevents robust
association with individual galaxies.

The sample of pairs resolved by IRAS (SSM) offers some insight into
the completeness of the IRAS identifications we make in the  BGK sample.
The BGK sample contains all pairs
in a magnitude limited redshift survey with projected separations
$< 50 h^{-1}$ kpc; the SSM pairs are restricted to systems separated by
fewer than 3 galaxy diameters. Interestingly,  the BGK sample
contains 4 pairs within
the 5.24 Jy limit of the SSM sample, but they are
too widely separated 
to meet the SSM diameter criterion. 
Table 5 lists these pairs: column (1) gives the J2000 coordinate of the BGK
galaxy closest to the IRAS source,
column (2) gives the IRAS B1950 identification, and columns (3-6) give the IRAS
fluxes. 

In most of the SSM pairs, the galaxies
are too faint
at B to be included in BGK. However 33 of the SSM pairs
lie within the 
limits defined by the selection of the BGK sample. Among these, two pairs are
missing from the BGK sample because the galaxies were not resolved in the
original Zwicky et al. (1961-1968) catalog on which the redshift survey was based.
Unresolved pairs at small
angular separation are a limiting factor for all samples of close pairs.  
All told, the BGK and SSM samples have 31 pairs in common 
and differ by a total of 6 pairs in the range of
complete overlap.

As we did for the BGK pairs, we also extracted 2MASS magnitudes for the SSM pairs.
We use these data only to show that the SSM pairs include more extreme
dusty objects than the BGK sample. 
The Appendix compares further basic IRAS properties of the BGK and SSM pairs.
More detailed comparison of these pair samples is beyond the scope of this
paper because of the complex selection of the SSM pairs and because we do
not have spectroscopy for them.

\subsection {Spectroscopy}

BGK and BGK2 describe the spectroscopic observations and data reduction in detail.
Here we briefly review the procedures we used.

We made observations of 502 galaxies with the FAST spectrograph at the
1.5 meter Tillinghast telescope on Mt. Hopkins. We observed each galaxy for
$\sim$ 10-20 minutes through a 3$^{\prime\prime}$-wide slit and used a 300 line/mm
grating to disperse the light over 4000-7000\AA.

Our spectra are representative of the ``central'' region of each galaxy.
The apertures extracted from the flat-fielded data range in length from
1.74 to 29.7$^{\prime\prime}$ corresponding to
0.25 - 13.7 h$^{-1}$ kpc with a mean of 2.4 h$^{-1}$ kpc (see Table 4).
The measured equivalent widths are the ratio of the flux in the line 
to the surrounding continuum corrected for Balmer absorption. 
We estimate the amount of absorption by taking the maximum absorption
equivalent width in H$\delta$ or H$\gamma$, and then adding that
equivalent width to both H$\alpha$ and H$\beta$.
Table 1
of BGK2 lists the EW(H$\alpha$) and H$\alpha$/H$\beta$ for the galaxies
in our sample. The typical error in the  EW(H$\alpha$) is 10\%; in the
Balmer decrement, the error is typically 20\%. 
Repeat measurements suggest that the error results largely
from uncertainty in the slit position.

We remove active galactic nuclei (AGNs) and ``ambiguous''
objects which may be AGN dominated from both
the BGK and NFGS samples by using the theoretical optical
classification scheme developed by Kewley et al. (2001). There are 3 such
objects in the NFGS (Kewley et al. 2002); there are 32 objects in BGK spectroscopic
sample. Because the number of these objects is small, removal has a negligible
effect on the analyses below; we remove the objects for consistency in our
focus on star formation. Among the 265 galaxies without spectra, we expect
about 17
AGN and/or ``ambiguous'' galaxies, a residual number too small to impact 
any trends in the dataset. In any case,  the mean J-H  of
the BGK AGN and ``ambiguous'' galaxies is 0.02 mag
bluer and their H-K is 0.02 mag redder than the 470 non-AGN galaxies in 
the spectroscopic sample.

\section {The Near Infrared Properties of  Galaxy Pairs}

Although the near infrared colors of galaxies span a narrow range, they
provide an interesting window on star-forming galaxies. In this section we
explore the infrared properties of the BGK sample.
Even though the BGK sample is B-selected, the sample is nearby
enough that the pairs can be resolved in 2MASS and {\it all} of the galaxies
are detected in 2MASS.

The near-infrared probes a complex combination of stellar population,
reddening, and gaseous and thermal dust emission. We use our own infrared photometry
and spectroscopy to probe these issues. For this exploration our photometry
has two advantages over 2MASS: (1) we extract photometry in the
aperture where we have spectroscopy and (2) we use the standard K filter
which extends to longer wavelength than the 2MASS K$_s$. The K filter is
more sensitive to thermal emission from 600 --- 1000K dust. 

\subsection {The Near Infrared Color-Color Diagram} 

The near infrared colors for a normal unreddened stellar population span a small
range in J-H and H-K$_{\rm s}$ (Aaronson 1977, Giuricin et al. 1993).
By collecting data
from two sets of close pairs selected either by optical morphology or by
projected separation and line-or-sight relative velocity, Giuricin et al. concluded that
``interacting''
galaxies display normal J-H colors and redder H-K colors than a ``normal'' galaxy population.
Making use of the existing L-band data, they also concluded that the redder H-K colors indicate
the presence of thermal emission from hot 600 --- 1000K dust. 
Here we compare the  J-H and H-K colors
for the BGK sample of close pairs with the NFGS sample of ``normal'' galaxies and
with the SSM sample of IRAS-selected pairs.

We have assembled data for a much larger sets of both pairs and normal 
galaxies than considered by Giuricin et al. (1993) and Aaronson (1977),
respectively. The errors in the infrared colors for galaxies in the
Giuricin et al. (1993) sample are typically 0.1 in J-H and 0.06 in H-K,
substantially exceeding the errors in our data. The sample considered by
Giuricin et al. (1993) is an exhaustive compilation of inhomogeneous
infrared data from the literature,
but the sample is not complete in any band. The pair sample we consider is
a complete B limited sample with near infrared photometry for all objects.
We examine the sample in the broader context of a similarly selected sample
of ``normal'' galaxies from the NFGS and the far infrared selected 
sample of SSM. In all cases the near infrared data are uniformly acquired
from 2MASS. 

\begin{figure}
\plotone{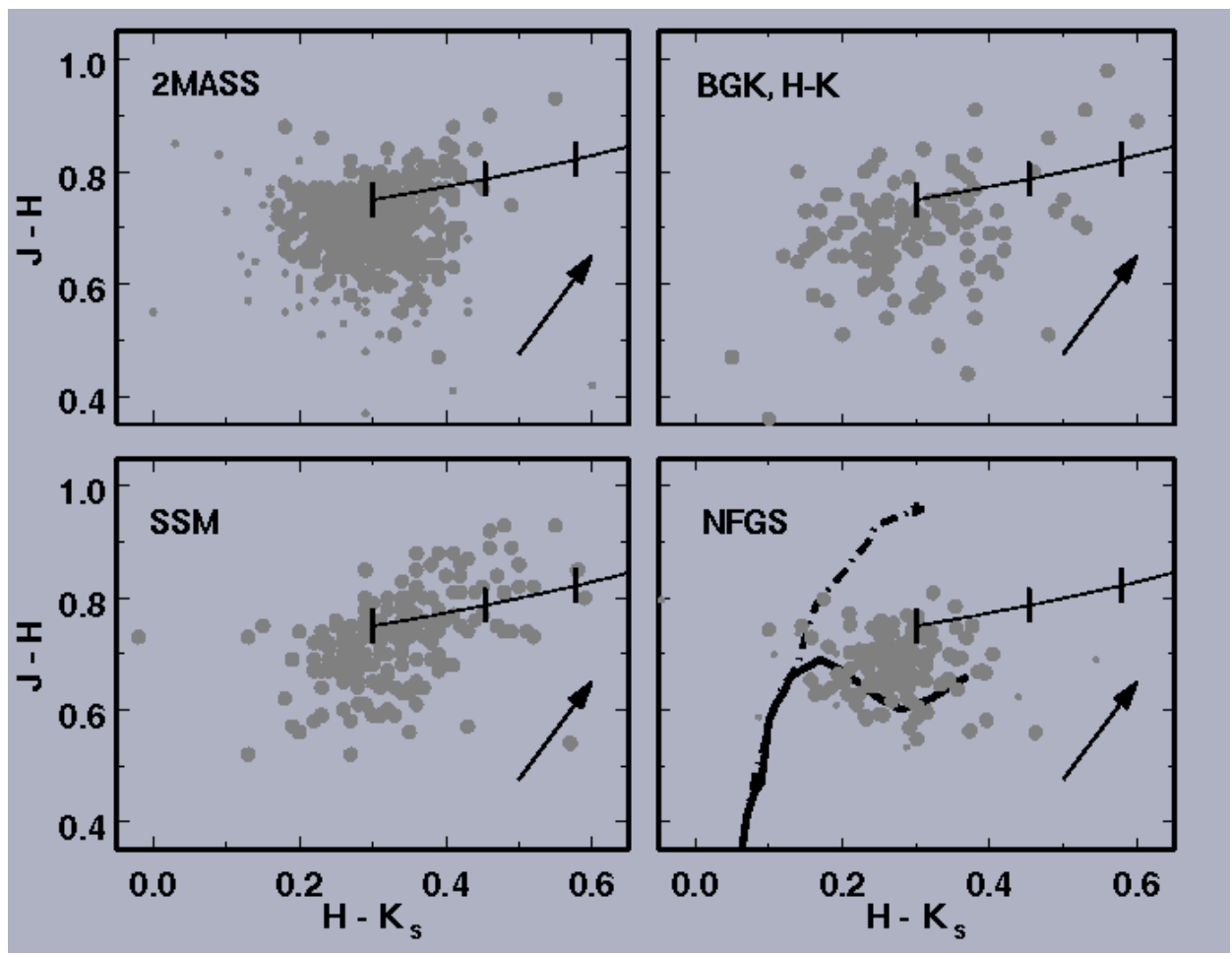}
\caption{Near infrared color-color diagrams for the BGK sample
2MASS data (upper left), the BGK small aperture JHK data (upper right,
this paper), SSM pair 2MASS data (lower left), NFGS ``normal'' galaxy 2MASS
data (lower right). Note the large spread of colors in the pair samples relative to the
NFGS. The small points denote galaxies with J $>13$; these objects tend to be
blue as a result of the original B-selected sample. In each panel the arrow
shows the reddening vector, the curve with ticks shows the
fractional contribution of 1000K dust at K; each tick marks a 10\%
increment. In the NFGS panel, the curves show the color of main sequence
stars (solid) and the giant branch (dot-dashed) from Bessel and Brett
(1988).}
\label{fig:ircolor}
\end{figure}

The panels of Fig. \ref{fig:ircolor} show color-color diagrams
for the 2MASS JHK$_s$ data (upper left) and for our JHK data (upper right) for the
individual galaxies in the BGK pairs. To set the BGK sample in a broader context,
we also  show 2MASS data for the individual galaxies in the SSM pairs
(lower left) and for the NFGS sample of ``normal'' galaxies. 
K-corrections for galaxies in these samples are small,
typically $\lesssim 0.03$ at K$_s$ and $\lesssim 0.01$ at J and H (Poggianti 1997). 
In the NFGS panel
(lower right), the
curves show how the colors of solar neighborhood
stars on  the main sequence (solid curve) and giant branch
(dot-dashed curve) bracket the
``normal'' galaxy colors \citep{bessell88}.

In all of the panels of Fig. \ref{fig:ircolor} the line
with ticks shows the contribution at K from thermal emission
by dust with a
temperature of 1000K; each tick marks a 10\% increase in the hot dust
contribution. The arrow indicates the reddening vector \citep{bessell88}.  

The NFGS sample of ``normal'' optically selected galaxies
spans the expected narrow range of near infrared colors.
K-S tests show that the probability that the J-H and H-K$_s$ distributions
for the BGK and NFGS samples are
drawn from the same parent distributions are  1.6$\times$10$^{-12}$ and
8.5$\times$10$^{-10}$, respectively. Both the mean J-H and H-K are slightly
redder in the BGK sample than in the NFGS and the spread in both colors is
larger, consistent with the results of Cutri and McAlary (1985). 
Giuricin et al. (1993) were unable to see the difference in the distribution
of J-H for pairs in their smaller sample. 

In our sample of pair galaxies the scatter in near infrared colors is
both toward red J-H and H-K and toward blue J-H and H-K than the
NFGS. As we discuss below this behavior results from the presence of
young central bursts of star formation which scatter the colors blueward and
from dustry bursts which scatter the colors redward. The original selection
of the sample at B and subsequent "observation" in the near
infrared favors detection of the blueward scatter relative to the
detection of dusty bursts.  Selection in a redder band favors the
redward scatter relative to the blueward scatter introduced by relatively
unobscured bursts; we demonstrate this point by examining the SSM
IRAS-selected pairs.

As expected, the SSM sample, selected at 60$\mu$m,  shows a notable extension toward red
H-K$_s$, much more impressive than in our B-selected dataset. Because we do not have complete spectroscopy for this sample, we
cannot remove the AGNs; however, it is improbable that all of the reddest
objects in H-K$_s$ are AGN. There are 16 known AGNs in the SSM sample
(Veilleux et al. 1995; Kewley et al. 2001; Corbett et al. 2003). These
AGN have mean near infrared colors J-H = 0.74$\pm$0.02 and H-K$_s$ = 0.33$\pm$0.01 consistent
with the overall mean for the sample. Only one of these AGN is a red outlier
in H-K$_s$. Based on the properties of the known AGN, we expect only 2-4 red outliers
in H-K$_s$ even if 30\% of the SSM galaxies are AGN. The well-populated
extension toward red H-K$_s$ is thus very probably a property of
star-forming objects.    

Without longer wavelength data, particularly 
at L, there is also an ambiguity in
the relative contributions of reddening and dust emission for 
star-forming objects with $0.4 <$ H-K$_s < 0.5$. 
However, even in the narrow K$_s$ filter, it is
essentially impossible to explain
the full extension of the colors of star-forming galaxies in the SSM sample
along the dust emission track in Fig. \ref{fig:ircolor}
by any other straightforward mechanism (Hunt et al. 2002). 

In our B selected pairs sample, the bluest galaxies are generally
faint in the near infrared; these objects also tend to have the lowest
intrinsic B-band luminosities in the pairs sample. Many of these objects
have J$> 13$ and thus appear as small points in
the color-color diagram. They tend to appear at blue J-H with a range
of H-K$_s$. It is interesting that the SSM sample which includes nearby 
intrinsically low luminosity objects also includes pair galaxies
with these bluer J-H colors. Overall, the
larger range of near infrared colors of pairs compared with the NFGS is
independent of the inclusion of these fainter objects although they
accentuate the difference in the expected sense.
Appendix A contains a more extensive comparison of the properties
of the SSM and BGK samples.

The upper right hand panel of Fig \ref{fig:ircolor} shows our JHK
slit aperture photometry 
for a subset of 160 BGK galaxies. More objects in this subset extend
toward redder J-H and H-K than in the 2MASS BGK
color-color diagram (upper left panel). There are two reasons for this extension.
Although color gradients are generally small in the infrared, some objects
are significantly redder in their central regions (Griersmith et al.
1982; Glass \& Moorwood 1985; Devereux 1989; Bushouse \& Werner 1990;
Carico et al. 1990; Jarrett 2000; Kewley et al. 2005).
The broader K filter is also more sensitive to hot dust emission which has a
negligible effect on J-H but produces a redder H-K. We discuss
these JHK colors in
more detail below.

\begin{figure}
\plotone{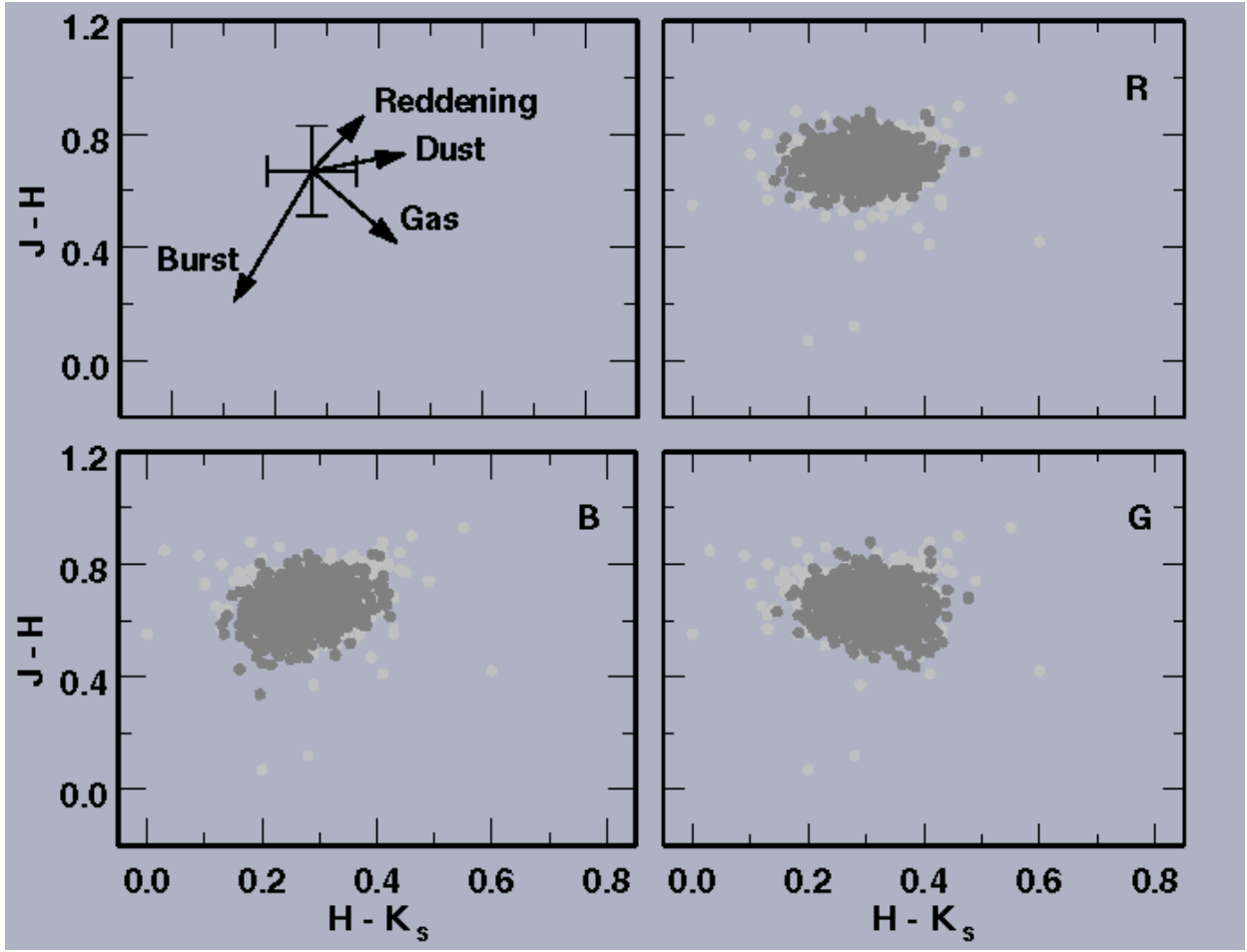}
\caption{Heuristic statistical model for the 2MASS BGK pairs color-color
diagram. The upper left panel is a schematic showing the physical
components of the model. The cross indicates the colors of a ``normal''
galaxy population. The arrows show the direction a galaxy moves
in this color space as a result of reddening, hot dust emission, gaseous emission, and young
bursts as indicated. In the other three panels gray points show the
data; black points show the model. In the upper right, the model (R,
Table 6) contains
only the scatter in colors and reddening; in the lower left we add bursts
of star formation
emission (B; Table 6) and on the lower right we add
gaseous emission (G; Table 6). Table 6 gives the
model parameters. In each panel, note the change in shape of the central mass of
points and the behavior of the outliers. The models account for
most of the observed spread of colors.}
\label{fig:irmodel}
\end{figure}

To understand the difference between the color-color diagrams for the BGK
pair sample and NFGS sample of normal galaxies,
we construct a simple model. The model is intended to account qualitatively
for the spread of colors rather than to explain details of individual cases.
The model has six ingredients: (i) an estimate of the mean color for a
``normal'' galaxy population, (ii) the spread in color which represents both
the intrinsic spread of a ``normal'' population and the measurement error,
(iii) reddening, (iv) 
emission from a young stellar population associated with a burst of star
formation, (v) gaseous emission from HII regions, and (vi) thermal reradiation from hot dust. 
The schematic in Fig. \ref{fig:irmodel} shows the normal range of near infrared colors (error
bar) along with arrows indicating the direction of the effect of the various
contributors to the near infrared colors. 

To reconstruct the distribution of colors for galaxies in the BGK sample,
we begin with the median colors of the NFGS:
(J-H)$_o$ = 0.67 and (H-K$_s$)$_o$ = 0.27.
The mean E$_{B-V}$ for the NFGS in our luminosity range
is  0.28$\pm$0.03. The median NFGS colors are completely consistent with earlier
studies (see e.g. Aaronson 1977; Giuricin et al. 1993). 

To account for
the spread of colors in a ``normal'' galaxy population and
for the measurement error in the colors, we first
add a dispersion of 0.05 mag in both colors, denoted by $\sigma_{J-H}$
and $\sigma_{H-K}$ in Table 6. With a typical 2MASS color error of 0.03 mag,
the intrisic dispersion in color we assume is 0.04 mag, consistent with
the results of Aaronson (1977).  
We select a random additional reddening by taking the absolute value of
a Gaussian with a dispersion of 0.020 in the
extinction at H-K$_s$, E$_{H-K_s}$. We denote this contribution
by $\sigma_R$ in Table 6. The extinction
at J-H,  E$_{J-H}$
= 1.95  E$_{H-K_s}$. The dispersion $\sigma_R$ corresponds to an {\it additional}
A$_V = 0.3$ (Bessell \& Brett 1988) and moves the colors redward along the
reddening vector.
Fig. \ref{fig:irmodel} (model R: upper right panel) shows that this simple prescription accounts
for most of the range  and for the redward shift in the
color-color diagram for the BGK pairs, but outliers remain particularly
toward the bluest J-H and reddest H-K$_s$. 

We can account statistically for most of the outliers by including 
contributions from  young stellar populations, free-free emission and dust.
Models B and G explore the effect of adding a burst of star formation and
gaseous emission respectively. Table 6 gives the parameters for these models;
column (1) gives the model designation, columns (2) and (3) list the fiducial
colors, (J-H)$_0$ and (H-K)$_0$, columns (4) and (5) give the ``normal'' spread of 
J-H and H-K colors, respectively, columns (6-9) give the contributions from
a young burst, hot dust, gaseous emission, and reddening, respectively.
The models are not unique; they are intended to demonstrate the impact of the
various contributions to the near infrared emission and to show that
reasonable parameters account for the observed spread of colors.

Model B in the lower left panel of Fig. \ref{fig:irmodel} shows 
model R with the addition of a burst contribution. In this example the burst
color is J-H = H-K$_s$ = 0, characteristic of A stars. For this demonstration,
we ignore the complexities of variation in the initial mass function,
metallicity, and spread in age and duration of the bursts.
We select the fractional contribution of the
``burst'' at H by taking the absolute value of a Gaussian with a
dispersion of 0.05; we denote this contribution by $\sigma_B$ in Table 6.
This choice is consistent with BGK2s optical analysis of burst strengths.
For the 40-60\% burst strengths most common in the BGK2 sample, a 5\% burst
contribution  at H corresponds to a roughly 50\% contribution at R for the
typical R-H$\sim$ 2.5 we measure in the slit aperture for the bluer objects.
Addition of a ``burst'' moves the colors blueward in both J-H and H-K$_s$.

Model G in the lower right panel of  Fig. \ref{fig:irmodel} shows model R
with the addition of  gaseous emission
with a color of J-H = 0 and H-K$_s$ = 0.7 (Campbell \& Terlevich 1984;
Whitelock 1985; Larios \& Phillips 2005)
The addition
of gaseous emission moves the colors blueward in 
J-H and  redward in H-K$_s$. We select the fractional contribution
of the gaseous emission at H by taking the absolute value of a Gaussian with
a dispersion of 0.08 and denote this contribution by $\sigma_G$ in Table 6. 

Fig. \ref{fig:irmodel} shows that
our heuristic model accounts for the colors of all but a few extreme outliers.
The outliers with blue H-K$_s$ and red J-H are fainter than J = 13 and may
have large errors in their colors. The two very blue objects in J-H may
be dominated by a strong burst and/or gaseous emission. The reddest
object in both J-H and H-K$_s$
requires at least a 10\% contribution from hot dust emission at K$_s$.

\begin{figure}
\plotone{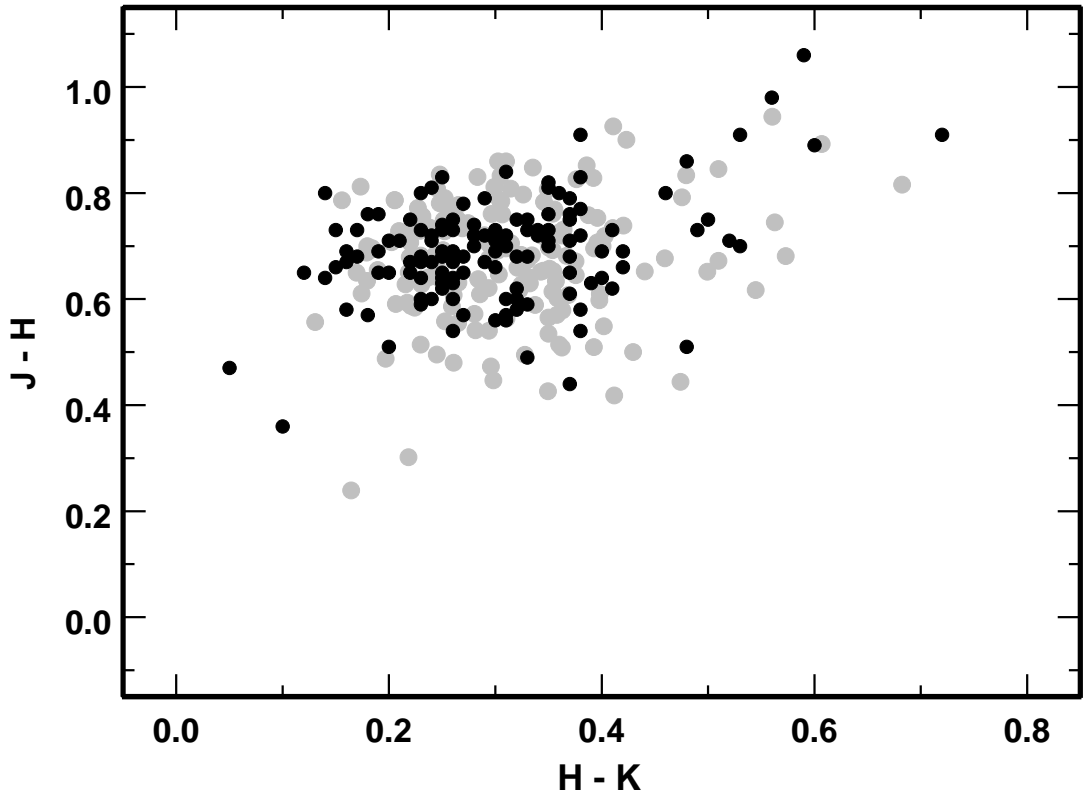}
\caption {Comparison of slit photometry colors for 160 BGK galaxies with the model
(parameters in Table 6; BGK). Gray points represent the data; black
points show the model. The K-band (as opposed to K$_s$ photometry)
is more sensitive to hot dust emission, 
necessary to account for the reddest objects in H-K.}
\label{fig:JHKmodel}
\end{figure}

For the BGK sample,  our small
aperture broader K band data  underscore the
necessity of accounting for emission from hot dust in modeling the
interaction. These conclusions are in accord with Giuricin et al. (1993).
In the model shown in Fig. \ref{fig:JHKmodel},
(J-H)$_o$ = 0.72; (H-K)$_o$ = 0.20, the error in J-H and H-K is 0.06 mag,
corresponding to a color error of 0.04 mag and an intrinsic spread in
color of 0.045 mag. We draw the
additional reddening  from a Gaussian with a dispersion of 0.06 in
H-K, the fractional contribution from gaseous emission at H is drawn from
a Gaussian with a dispersion of 0.04, the fractional ``burst'' contribution
at H is drawn from a Gaussian with a dispersion of 0.08. To match the
extension toward the reddest H-K, we introduce a contribution from dust emission. The fractional dust
emission
contribution at H is drawn from a Gaussian with a dispersion of 0.008
and denoted by $\sigma_D$ in Table 6. At 
K, this dispersion is 0.09. Thus the dust contribution for the reddest
objects in H-K is 10-20\% at K. 

\begin{figure}
\plotone{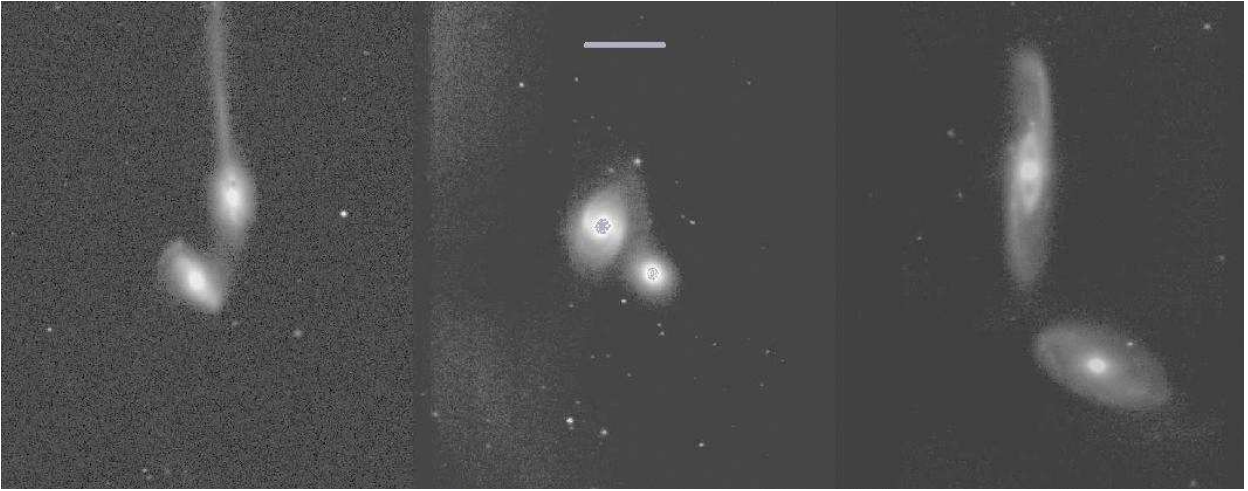}
\caption {Near-infrared images for three galaxy pairs. 
The images have North up and East to the left
and scales of 0.34 arcsec pixel$^{-1}$. The white 
bar in the central image has a length of 10 arcsec. 
(a) Left panel: J-band image of 124344.0+310017.9
(H-K = 0.60) and 124345.3+305943.8 (H-K = 0.37).
The northern component has a dust lane on JHK
images.
(b) Middle panel: K-band image of 152418.9+415040.7
(H-K = 0.28) and 152420.9+415059.5 (H-K = 0.72).
The redder eastern component has a tidal tail not
visible on the DSS image.
(c) Right panel: K-band image of 113704.8+321108.9
(H-K = 0.82) and 113707.1+321227.2 (H-K = 0.22).
The southern component is a red AGN which we exclude from the sample;
the blue northern
component, retained in the sample, has a bright ring of star formation. The 30$^{\prime\prime}$
white bar gives the image scale.}

\label{fig:images}
\end{figure}

Fig. \ref{fig:images} shows images of a few of the
objects in our infrared 
small aperture photometry sample with the reddest
central H-K colors. For the red objects in
the left and central panels, there is no spectroscopic evidence of AGN
activity. We exclude the AGN in
the righthand panel from the sample; its blue companion
has an EW(H$\alpha$) = 37\AA.

The Infrared Space Observatory (ISO) provided the first suggestion of hot
600-1000 K dust in star-forming galaxies (Helou et al. 2000). Subsequent
JHKL$^{\prime}$ observations demonstrate that some actively star-forming galaxies have 
K-L$^{\prime} \geq 1$, consistent with hot dust emission (Hunt et al. 2002).
Hunt et al. (2002) suggest that
the hot dust is associated with the intense far ultraviolet radiation field in
compact ($\leq$ 100 pc) regions of active star formation which might arise naturally
from galaxy-galaxy interactions and mergers. In their sample of 26 galaxies, they
were unable to detect any correlation between the presence of hot dust and the much
cooler dust.  

For the BGK pairs the ratio of 60 $\mu$m to 100 $\mu$m IRAS fluxes implies
emission from dust at temperatures in the range 20-70K assuming
a blackbody spectrum (Beichmann et al. 1985).
We do not have enough pairs with both small aperture photometry and IRAS
detections to make a meaningful comparison of H-K with
L$_{FIR}$/L$_H$; we thus use the 
much larger sample of 2MASS H-K$_s$ data to see whether there is any
relationship between the presence of hot dust and far infrared 
emission from cooler dust.
 
\begin{figure}
\plotone{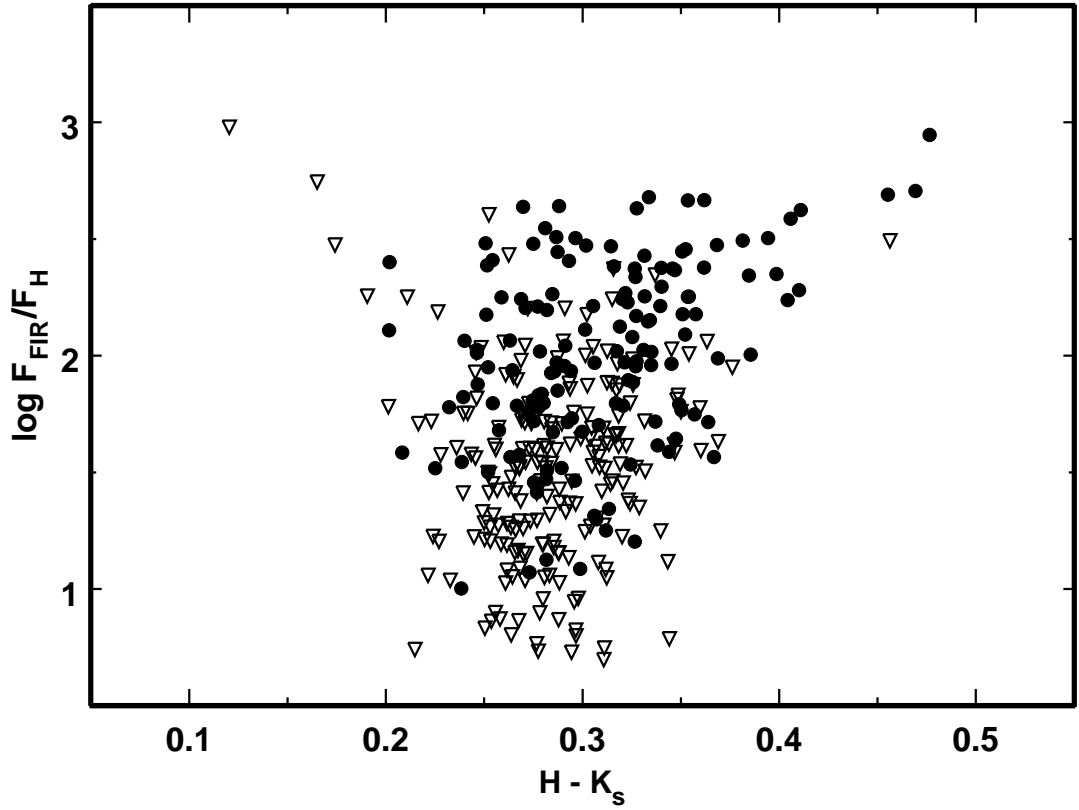}
\caption {L$_{FIR}$/L$_H$ for the BGK pairs. The solid points are
detections; the open triangles are upper limits. }
\label{fig:FIRH-K}
\end{figure}

Fig. \ref{fig:FIRH-K} shows the distribution
of the normalized far-infrared luminosity
L$_{FIR}$/L$_H$ as a function of H-K$_s$ for the BGK pairs.
We define the far infrared flux in the standard way
(Helou et al. 1988; Sanders \& Mirabel 1996):
$$ {\rm F_{FIR}} = 1.26 \times 10^{-14} \{2.58 f_{60} + f_{100}\} {\rm W m^{-2}} $$

\noindent where f$_{60}$ and f$_{100}$ are the 60$\mu$m and 100$\mu$m fluxes
respectively. If a pair is resolved we sum the fluxes for the components;
for each pair the luminosity we plot represents the total for the pair
and the H-K$_s$ color is the appropriately weighted ``pair'' color.

Fig.  \ref{fig:FIRH-K}  shows  L$_{FIR}$/L$_H$ as a function of H-K$_s$ from 2MASS for
the BGK pairs for both IRAS detections and upper limits. 
L$_{FIR}$ is proportional to the star formation rate (Kennicutt 1988; Calzetti et al.
2000;
Charlot et al. 2002; Kewley et al. 2002) and L$_H$ is a measure of the stellar mass.
Thus  L$_{FIR}$/L$_H$ provides a measure of the normalized or specific star formation rate.
Fig. \ref{fig:FIRH-K} shows that the reddest 
H-K$_{\rm s}$ colors correspond to  relatively large F$_{FIR}$/F$_H$.
Using the ASURV package (Lavalley et al. 1992), the Spearman rank
probability of
no correlation including the upper limits
in Fig.  \ref{fig:FIRH-K}  is $ < 10^{-4}$.
The increasing scatter in H-K$_s$ with increasing F$_{FIR}$/F$_H$ is also a
striking feature of Fig.  \ref{fig:FIRH-K}. At the bluest H-K$_s$, all of the points with
large F$_{FIR}$/F$_H$ are upper limits (open triangles); nearly all of the
reddest H-K$_s$ are detections at large  F$_{FIR}$/F$_H$. The data suggest that
the reddest galaxies in H-K$_s$ contain dust emitting over a wide temperature
range. These data indicate that, as suggested by Hunt et al. (2002), some galaxies
which undergo strong central bursts of star formation and thus have
substantial specific star formation rates contain compact star-forming
complexes with very red H-K$_s$.  These regions may be similar to the dusty super-star clusters
found in interacting galaxies and ULIRGs (e.g. 
Gallagher \& Smith 1999; Bekki \& Couch 2001; Keel \& Borne 2003; Kassin et al. 2003) 

Emission from hot dust may also be associated with AGN activity.
We have removed the spectroscopically identifiable AGN from our sample. However,
we show in the next section that the reddest objects in H-K have weak or even
undetectable H$\alpha$ emission. It is possible that  AGN activity is
hidden. Because AGNs are rare in the large sample we can classify spectroscopically,
we favor vigorous star formation as the explanation of the reddest H-K colors.
Hard x-ray observations with {\it Chandra} and mid infrared imaging with {\it Spitzer} would be useful in better identifying AGNs (Alonso-Herrero et al.
2006).

It is frustrating that in Fig.  \ref{fig:FIRH-K}, the poor resolution of IRAS prevents
plotting quantities for individual galaxies that can be readily identified
with points in the color-color diagrams. Clearer tests of the underlying
physics which dominates the color-color plots requires the resolution of
{\it Spitzer}. Hopefully large objectively selected samples of nearby pairs
with {\it Spitzer} observations will soon be forthcoming.

\subsection {Spectroscopy and Infrared Colors}

\begin{figure}
\plotone{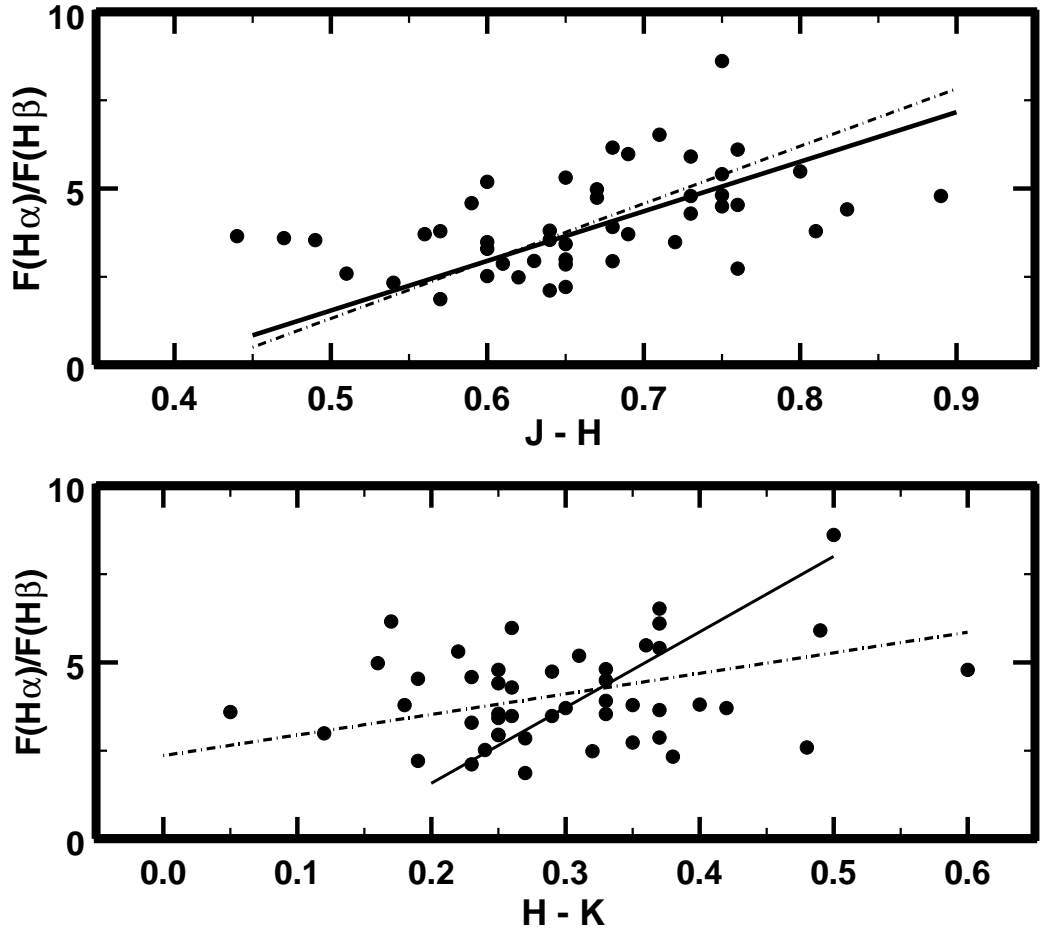}
\caption{Balmer decrement for 45 BGK galaxies as a function of J-H (top) and H-K (bottom).
The colors are measured in the spectroscopic aperture. The solid
line shows the Galactic reddening law;  the dot-dashed lines show the
best fit slope. For J-H the fitted slope is essentially the same as the
Galactic reddening law; for H-K the slopes differ because hot dust emission
affects the H-K color.}
\label{fig:Balmer}
\end{figure}

We next use our spectroscopy to elucidate further the physics which underlies
the color-color diagrams. We combine the spectroscopy with our JHK
photometry in the spectroscopic aperture.    
Fig. \ref{fig:Balmer} shows the relationship between the Balmer
decrement and the J-H color (see Moorwood et al. (1987) for the first
demonstration of this particular correlation but without a fitted slope). 
With the exception
of a few outliers, it is
remarkable that these objects, some of which contain young or
heavily reddened bursts of star
formation, have essentially the same reddening
law as the Galaxy.  The Spearman rank probability of no
correlation is 10$^{-4}$ and the best fit slope (16.2$\pm$1.9,
$\chi^2/{\rm dof} = 2.1$; dot-dashed
line) is close to the
Galactic reddening law (14.0; solid line; Bessel \& Brett 1988).   

\begin{figure}
\plotone{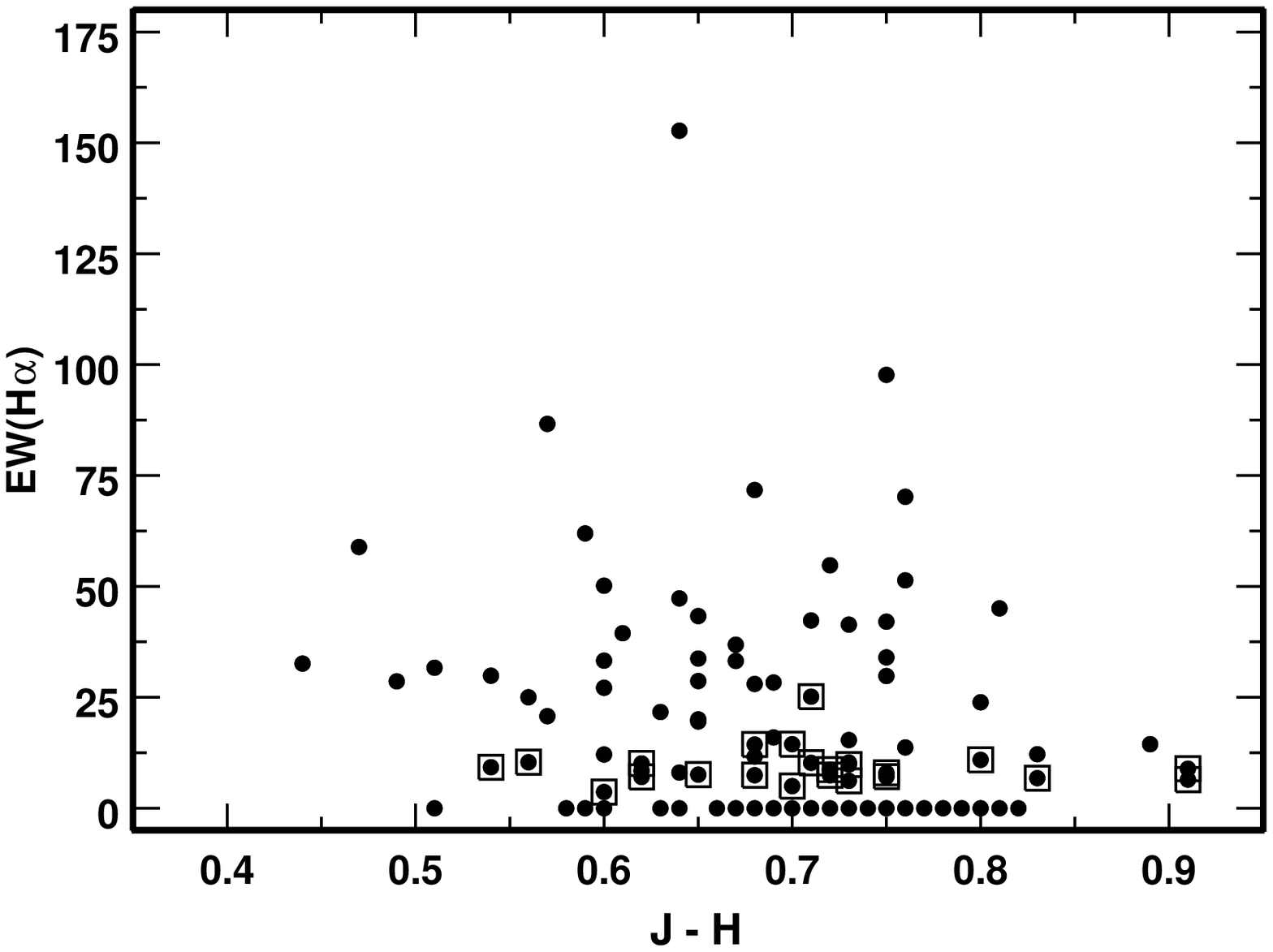}
\caption{Equivalent width of H$\alpha$ as a function of J-H for 151 BGK galaxies. The color
is measured in the spectroscopic aperture. Boxes denote galaxies with non-zero
EW(H$\alpha$) but undetectable H$\beta$.}
\label{fig:alphaJ-H}
\end{figure}

A large EW(H$\alpha$) indicates a strong burst of star formation. BGK2
conclude that these bursts can account for 40-60\% of the galaxy light at R. 
These substantial bursts should result in bluer than normal near infrared
colors. Fig. \ref{fig:alphaJ-H} shows the J-H color as a function of the 
EW(H$\alpha$).  
Indeed at the very bluest colors (J-H $\lesssim$ 0.57), 55\% of the galaxies 
have EW(H$\alpha$) $\gtrsim$ 25 \AA~and 9\% have EW(H$\alpha$) $\lesssim$ 
10 \AA. At J-H $\gtrsim$ 0.78, 11\% of the galaxies have EW(H$\alpha$) 
$\gtrsim$ 25 \AA~and 56\% have EW(H$\alpha$) $\lesssim$ 10 \AA.  The Spearman
rank probability of no correlation is 6.8$\times$10$^{-5}$. 
The EW(H$\alpha$) = 0 objects have a major impact on the Spearman rank
probability; eliminating them gives a probability of 0.11 of no correlation.

Fig. \ref{fig:Balmer} shows that the Balmer decrement 
is less well correlated with H-K; the Spearman rank probability
of no correlation is 0.26. We see little correlation
here because 
dust emission affects H-K but not J-H
(see the schematic in Fig. \ref{fig:ircolor} and the model in Fig.
\ref{fig:irmodel} which requires
dust emission to account for the data).
The dot-dashed line line indicates the best fit slope 5.8$\pm$0.4,
with an abysmal $\chi^2/{\rm dof} = 41$. The solid line shows the slope of the standard
Galactic reddening law (21.4; Bessell \& Brett 1988).

\begin{figure} 
\plotone{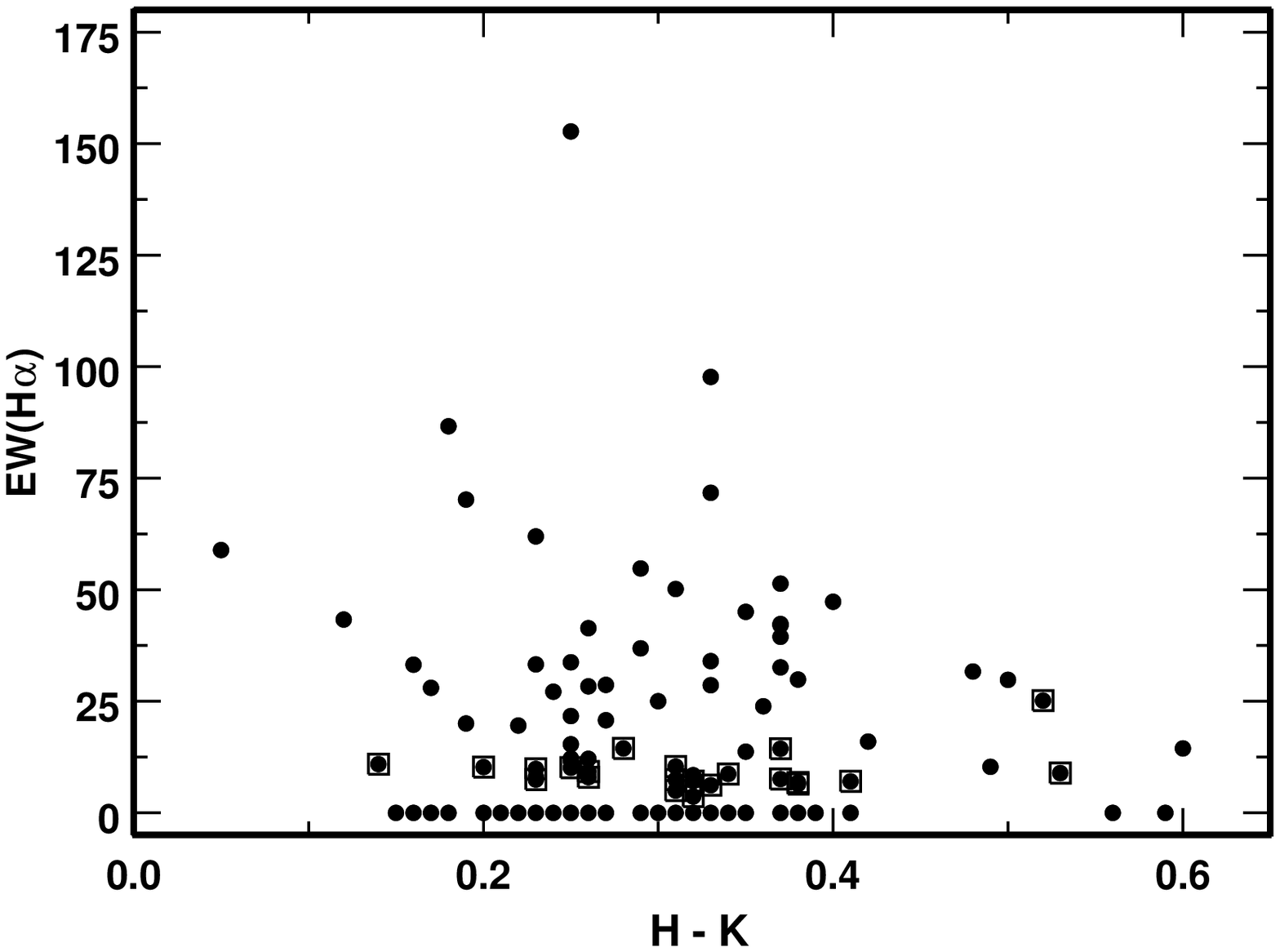}
\caption{Equivalent width of H$\alpha$ as a function of H-K for 151 BGK galaxies. The color is
measured in the spectroscopic aperture. Boxes denote galaxies with non-zero
EW(H$\alpha$) but undetectable H$\beta$.}
\label{fig:alphaH-K}
\end{figure}

Fig. \ref{fig:alphaH-K} shows the relationship between EW(H$\alpha$) and H-K.
The apparent correlation is not significant; the Spearman rank probability 
of no correlation is 0.65. 
The largest EW(H$\alpha$) occur at bluer H-K$\lesssim 0.35$; 10\% of galaxies
with H-K$\lesssim 0.35$ have EW(H$\alpha$) $\gtrsim$ 50 \AA. None of the
reddest galaxies (H-K $\gtrsim$ 0.45) have EW(H$\alpha$) $\gtrsim$ 50 \AA.
Although most of the reddest galaxies have EW(H$\alpha$) $\lesssim$ 10 \AA,
there are also objects at blue H-K with EW(H$\alpha$) = 0.

In Figs. \ref{fig:alphaJ-H} and \ref{fig:alphaH-K}, boxed symbols indicate
galaxies with non-zero EW(H$\alpha$) but without a measurable
F(H$\alpha$)/F(H$\beta$). Of the five objects with H-K $> 0.5$ (these objects are
the most likely to contain dusty compact regions of intense star formation) 
two have EW(H$\alpha$) = 0, two have immeasurable Balmer decrements, 
and one has F(H$\alpha$)/F(H$\beta$)$\sim$ 5. All of these objects
are in IRAS-detected systems  with L$_{FIR}$/L$_H$ $\gtrsim$ 1.8. 
The member galaxies 
thus may contain dust enshrouded bursts.  Accounting for these ``hidden'' bursts
is important for a full picture of tidally triggered star formation and
for calculations of the star formation rate density throughout the universe.

\section {The Infrared and Triggered Star Formation}

In this section we investigate the relationship between the infrared
properties of the BGK pairs and the projected separation of the pair, a measure
of the interaction. An impressive range of  
star-formation indicators are correlated with the projected separation
including the EW(H$\alpha$) (BGK, BGK2), the star formation birth rate parameter for pairs in the 2dF survey
(Lambas et al. 2003; Alonso et al. 2004), the mean specific star
formation rate derived from Sloan Digital Sky Survey (SDSS)
H$\alpha$ and z-band luminosities (Nikolic et al. 2004),  the concentration index
measured in the r-band as a measure of the presence of nuclear bursts of star formation
(Nikolic et al. 2004),  and metallicity as an indicator of gas infall in the BGK pairs
(Kewley et al. 2006).

\subsection { Far Infrared  Specific Star Formation Rates}

\begin{figure}
\plotone{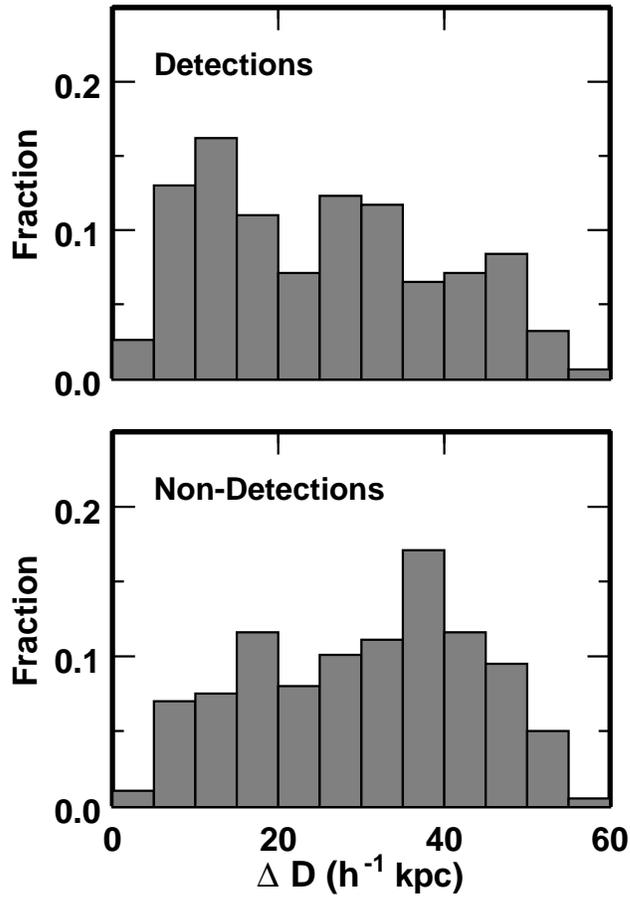}
\caption {Normalized distribution of projected separations for IRAS-detected
BGK pairs (top) and undetected pairs(bottom)}
\label{fig:IRASdet}
\end{figure}

Fig. \ref{fig:IRASdet} shows the normalized distribution of
projected separations for pairs
in the BGK sample with and without IRAS detections. 
The KS probability that the two distributions
are drawn from the same population is 3.3$\times$10$^{-4}$. The pairs detected
in IRAS have a median separation of 20 $h^{-1}$ kpc; the median for
the undetected
pairs is 32 $h^{-1}$ kpc. These distributions suggest that pairs which are
most probably close together in space have greater specific star formation
rates. 

\begin{figure}
\plotone{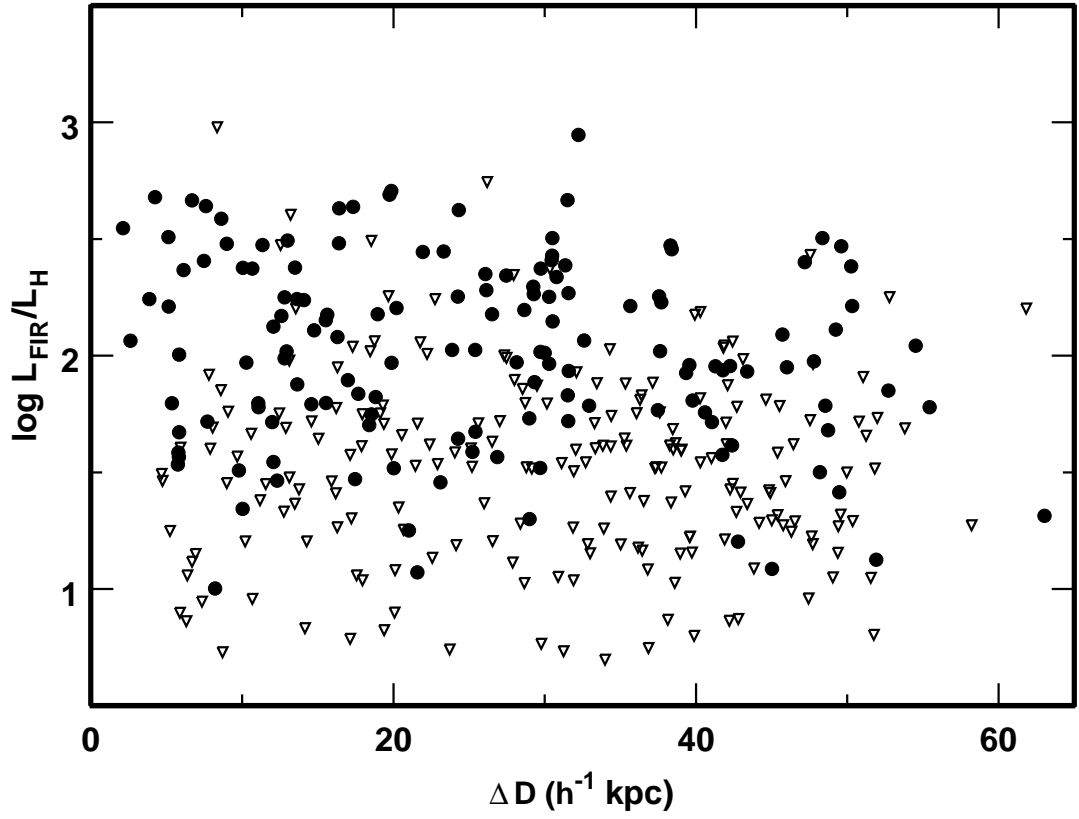}
\caption {L$_{FIR}$/L$_H$ as a function of separation for the BGK pairs.
Solid dots are detections, open triangles are upper limits. Note that
L$_{FIR}$/L$_H$ refers to individual pairs, not individual galaxies.}
\label{fig:IRASsep}
\end{figure}

To explore the connection between specific star formation rate and 
$\Delta{D}$ further, we calculate log(F$_{FIR}$/F$_H$) (solid dots) or
the upper limit on
log (F$_{FIR}$/F$_H$) (open triangles) for each BGK pair. Fig.
\ref{fig:IRASsep} shows the distribution of
log (F$_{FIR}$/F$_H$), a proxy for the specific star formation 
rate, as a function of $\Delta{\rm D}$, for pairs with 
$\Delta{\rm D} < 60$ h$^{-1}$ kpc.  

To examine the correlation between specific star formation rate and
$\Delta{\rm D}$ we use the ASURV package
(Lavalley et al. 1992) which treats the upper limits.
The Spearman rank probability of no correlation is $ < 10^{-4}$.
Smaller $\Delta{\rm D}$ favors larger log (F$_{FIR}$/F$_H$). 

\begin{figure}
\plotone{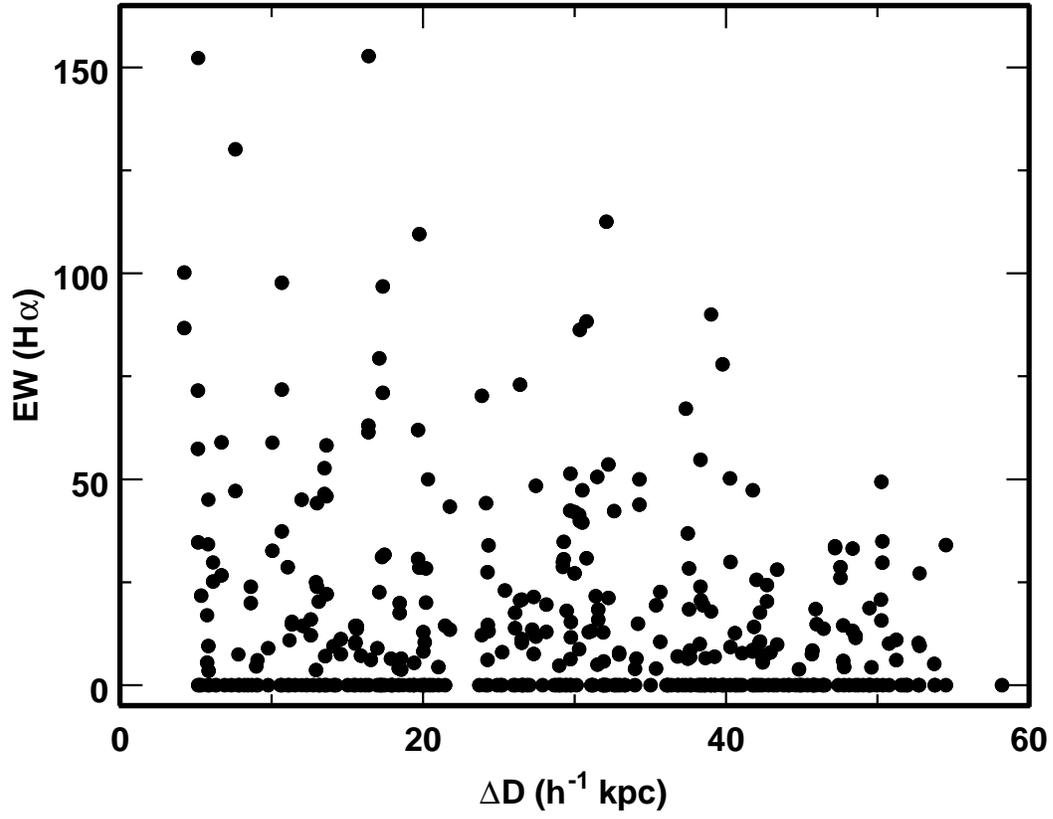}
\caption {EW(H$\alpha$) as a function of projected separation for BGK pairs.
Each point represents an individual galaxy.}
\label{fig:alphasep}
\end{figure}

For comparison with Fig. \ref{fig:IRASsep}, Fig. \ref{fig:alphasep}
shows the EW(H$\alpha$) as a
function of $\Delta{\rm D}$ for the entire BGK sample. In contrast
with Fig. \ref{fig:IRASsep} where the IRAS resolution limits us to plotting fluxes for
systems, we plot EW(H$\alpha$) for individual galaxies in Fig.
\ref{fig:alphasep}.

Nikolic et al. (2004) discuss the normalized FIR star formation rate as a function of
separation for a subset of their SDSS pairs. They consider only IRAS detections and
discuss the inherent bias in that approach. They conclude that the normalized star
formation rates calculated from H$\alpha$ emission or from FIR fluxes decrease with
increasing projected separation in agreement with our conclusion derived from both
detections and upper limits. A serious limitation on comparison of the spectroscopic and
FIR indicators of the star formation rate is that our spectroscopic measurements are
limited to a central aperture; the IRAS fluxes rarely even resolve the pair.

BGK and BGK2 plot
the relation in Fig. \ref{fig:alphasep} for subsets selected by local density.
Like Nikolic et al. (2004) and other investigators, we make no density
selection here. 
The Spearman rank probability that the data are uncorrelated
is  $ < 10^{-4}$, consistent with the result based on the IRAS fluxes and upper
limits for the same sample. 

The data provide strong evidence for a relationship between the normalized
star formation rate and the projected separation of the pair regardless of
the measure we use for the star formation rate. It would be fascinating to
see a similar plot for galaxies in pairs with complete spectroscopy observed
with {\it Spitzer} so that a more direct comparison could be made between the
star formation rate indicators for a large objectively selected pair sample.

\subsection {Near Infrared Colors}

The situation in the near infrared is complex. Dust emission produces a
redder H - K$_{\rm s}$ but leaves J-H essentially unchanged. Reddening produces 
generally redder near infrared colors and a burst of star formation
produces bluer colors. Gaseous emission produces bluer J-H and redder H-K.
Because of the multiplicity of sources of near infrared emission
and because the full range of near infrared colors of
stellar populations is small, the near infrared colors alone are not useful
for estimating the strength of a burst of star formation.

The near infrared
color-color diagram (Fig. \ref{fig:ircolor}) shows that the distribution
of near infrared colors of close
pairs differs from a sample of ``normal'' galaxies.  For a more complete 
model, L-band and resolved
60 $\mu$m photometry are necessary to separate the effects of reddening from the effects
of dust emission. Nonetheless, the near infrared colors
indicate a possibly interesting
dependence on the projected separation of the pair. 

\begin{figure}
\plotone{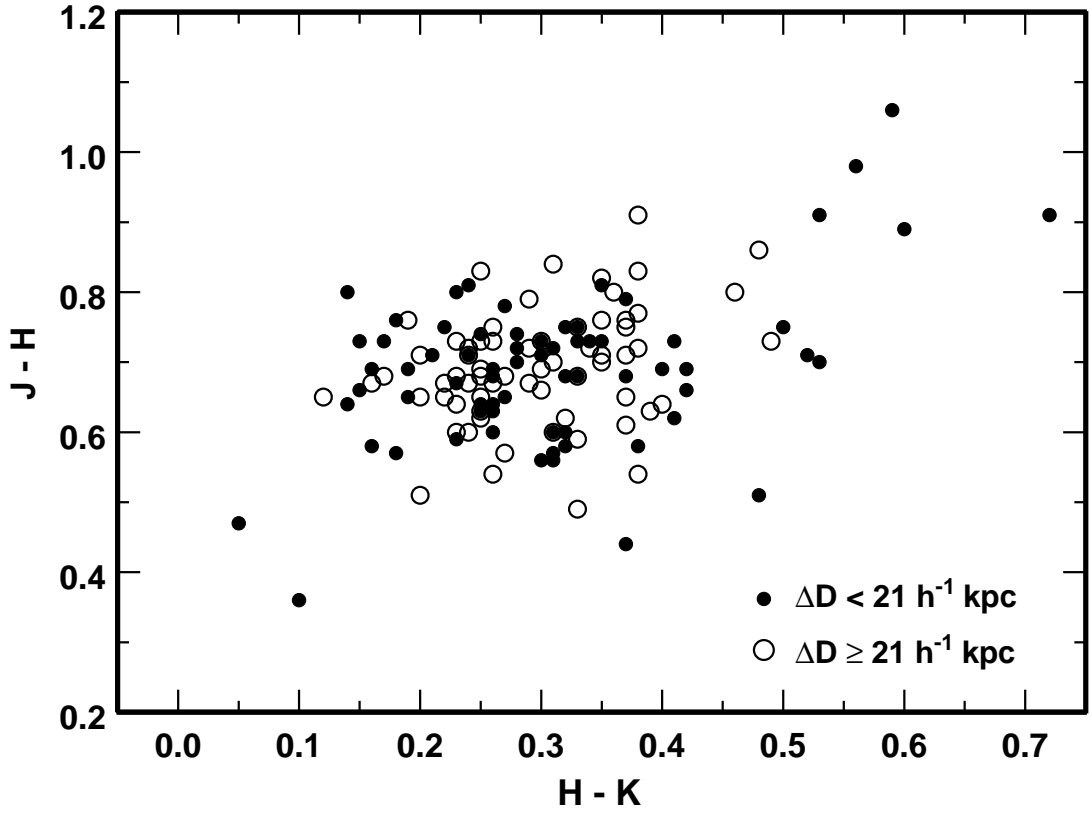}
\caption{Small aperture photometry near infrared color-color diagram for
160 galaxies in BGK pairs with projected separations 
$\Delta{D} < 21 h^{-1}$ kpc (solid dots) and with projected
separation $\Delta{D} \geq 21 h^{-1}$ kpc (open circles). }
\label{fig:ircolorsep}
\end{figure}

Fig. \ref{fig:ircolorsep} shows the near infrared color color diagrams for the
slit aperture JHK sample
divided at the median separation, $\Delta {\rm D} = 21 h^{-1}$ kpc.
There are 80 galaxies in each of the two subsamples. The
two subsamples appear different to the eye in the sense that the
spread in colors, particularly H-K, is larger at smaller separation;
all of the bluest and reddest objects at H-K are at smaller 
projected separations.
There
are outliers in H-K toward the blue (indicating central bursts of star
formation) and toward the red (indicating dusty bursts).
It is interesting that the outliers in the small $\Delta{D}$ sample are
in the directions indicated in the schematic of Fig. \ref{fig:irmodel}
for young burst and  dust emission contributions. All of the 
candidate dust enshrouded bursts 
with H-K $>$ 0.5 are in pairs with small projected separation. 
However, a 2D KS
test shows the samples have a 10\% probability of being drawn from the
same underlying distribution.

Fig. \ref{fig:ircolorsep} suggests that the range of colors
may be greater for
tighter pairs, but the strength of the conclusion may be limited by the
sample size. The 2MASS data also yield
insignificant differences in color distribution with projected
separation, but the 2MASS aperture is large and the 2MASS K$_s$ is
less sensitive to the effects of dust.
Exploration of a larger sample of small aperture data
would be worthwhile especially with the addition of L-band data.
There is potential for detecting both the impact of
dusty bursts and blue bulges in this approach.

\section {Conclusion}

We use the BGK sample of pairs of galaxies selected from 
the complete CfA2 redshift 
survey to examine near and far infrared photometry for clues to the
nature of the galaxy-galaxy interaction. The depth of the CfA2 survey is
well-matched to 2MASS and all of the nearly 800 galaxies in the sample 
have 2MASS photometry. The sample of close pairs we analyze is much
larger than those considered in previous investigations of the infrared
properties of close pairs.

We use a combination of 2MASS photometry, deep JHK photometry in small
apertures, spectroscopy, and IRAS data to explore the infrared as a probe of
triggered bursts of star formation. We find:

\begin{enumerate}
\item The distribution of J-H, H-K$_s$, and H-K colors of pairs 
is broader than the corresponding distribution for a sample of ``normal''
galaxies. We interpret this difference as evidence for bursts of star
formation which produce an extension toward bluer colors and for a
combination of reddened and/or dusty bursts which produce an extension
toward redder colors, particularly in H-K. In the color-color diagram the
reddest objects in H-K are also red in J-H. The colors of these objects follow
a track for
thermal emission from 600-1000K dust. The reddest  H-K colors require emission
from hot dust.

\item A statistical model including emission from a young stellar population,
gaseous emission from HII regions, and emission from hot dust explains the outliers
in infrared color-color diagrams. This model shows that triggered central bursts
affect the near infrared colors. The effects are complex because of the
multiplicity of emission processes important in the near infrared.

\item We use our spectroscopy and small aperture photometry to show that 
the central J-H  colors and Balmer decrements are consistent with the Galactic
reddening law. However H-K colors are essentially uncorrelated with the Balmer
decrement as a result of emission from hot dust.

\item We identify a set of objects with central H-K $\geq$ 0.45, with
F$_{FIR}$/F$_H \gtrsim 1.8$, and with small EW(H$\alpha$). We argue that these objects harbor compact
dust enshrouded bursts possibly similar to the super-star clusters identified in
well-studied interacting galaxies. These objects may be examples of ``hidden'' bursts
or, possibly, ``hidden'' AGN. Their presence supports the
contention that H$\alpha$ surveys
underestimate the volume averaged star formation rate in the nearby universe.

\item We examine measures of the specific star formation rate as a function of the
projected separation of the pairs in our sample. Both spectroscopic and
far infrared photometric measures show that the specific star formation rate
decreases with increasing projected separation. The poor spatial resolution of IRAS
prevents direct comparison of these measures.  

\item  Examination of the near infrared color-color diagram as a function of projected
separation shows that all of the outliers indicative of a central burst
of star formation lie in pairs at small separation. Although our sample is too small
to make the case for a statistically significant dependence of
the near infrared color-color diagram on projected separation, the
data suggest that larger samples of deep, small aperture near infrared 
photometry
would be a basis for identifying both the very blue and very red central
bursts. An increase in the dispersion might be expected
as a result of triggered bursts which produce blue central regions in the
absence of
dust and very red central regions in the presence of dust.

\end{enumerate}

Taken together these results indicate that a larger sample of pairs with
complete spectroscopy and with small aperture near infrared photometry and
resolved mid-infrared photometry from {\it Spitzer} would be an important
dataset for isolating the effects of young bursts, reddening, and thermal
dust emission over the course of galaxy-galaxy interactions and
subsequent mergers.

\acknowledgments
We thank Jason Surace for providing early versions of the computer readable tables
from SSM. We thank the anonymous referee for comments which prompted
several important clarifications.
This paper uses data from the {\it Infrared Astronomical Satellite,} 
a joint project of the Netherlands Agency for Aerospace Programs
(NIVR), the US National Aeronautics and Space Administration (NASA)
and the UK Science and Engineering Research Council (SERC).
This publication makes use of data products from the Two Micron 
All Sky Survey, which is a joint project of the University of 
Massachusetts and the Infrared Processing and Analysis Center/California 
Institute of Technology, funded by the National Aeronautics and 
Space Administration and the National Science Foundation.
This research has made use of the NASA/IPAC Infrared Science Archive, 
which is operated by the Jet Propulsion Laboratory, California 
Institute of Technology, under contract with the National 
Aeronautics and Space Administration.

{\it Facilities}: {Whipple Observatory 1.5-meter (FAST); Whipple
Observatory 1.2-meter (STELIRCAM); KPNO 2.1-meter (SQIID and ONIS)}
\appendix
\section{Appendix}
\subsection { IRAS Properties of the BGK and SSM Samples}

\begin {figure}
\plotone{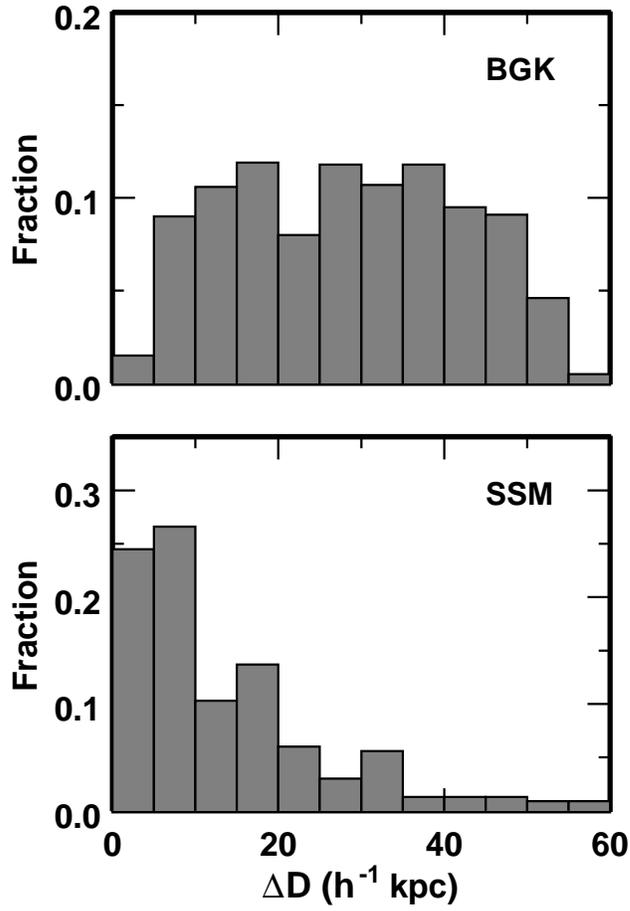}
\caption{Normalized distributions of projected pairwise separations
for the BGK and SSM samples.}
\label{fig:BGKSSMsep}
\end{figure}

Differences in the selection of samples of close pairs can have a marked
effect on the global properties of the ensemble of galaxies. The BGK and SSM
samples are an interesting case of selection in very different wavelength
ranges and with different separation criteria.
The BGK sample is selected 
from a complete magnitude limited redshift survey based on the Zwicky et al.
(1961-1968) catalog. As a result of the construction of the Zwicky
catalog, the BGK sample is deficient in pairs separated by 
$\leq5 h^{-1}$ kpc.  The SSM sample
is flux limited at 60 $\mu$m. The restrictive criterion of projected pairwise
separation to three average projected galaxy diameters biases the sample against widely
separated pairs.  Fig. \ref{fig:BGKSSMsep} shows the difference in the distribution of 
projected pairwise separations for the two samples.

\begin{figure}
\plotone{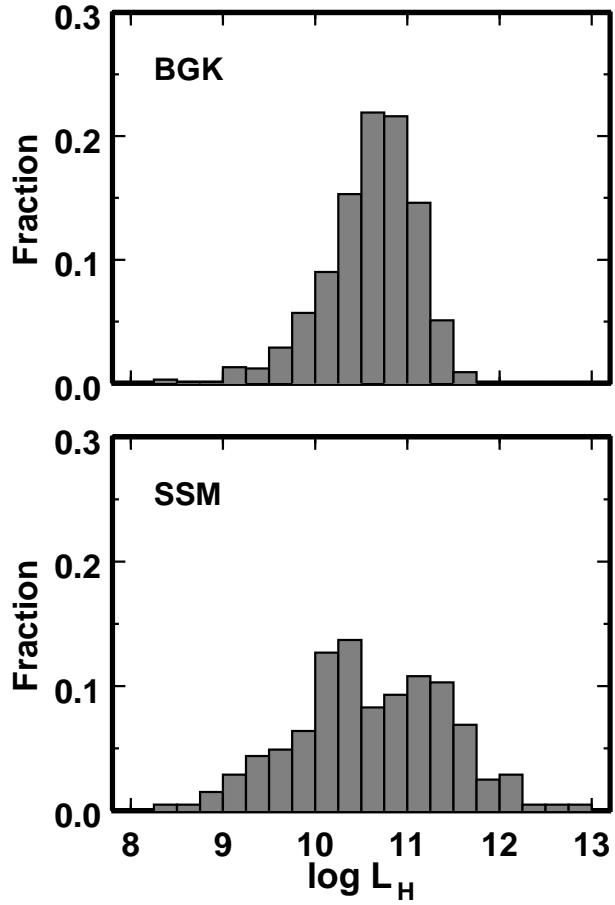}
\caption{Normalized distributions of L$_H$ for {\it individual galaxies}
in the BGK and SSM pair samples. Note
the broader distribution of SSM luminosities.}
\label{fig:LH}
\end{figure}

Fig. \ref{fig:LH} shows that the SSM sample spans a somewhat broader
range of H band luminosity, and the BGK sample is
more narrowly peaked. The  mean $cz$ of the SSM sample is
4287 km s$^{-1}$; for BGK,
it is 5852 km s$^{-1}$ explaining the concentration of the BGK distribution
toward greater H luminosities. Very
low luminosity objects appear predominantly in the 
SSM sample because BGK have a lower redshift cutoff of 2300 km s$^{-1}$ which
eliminates these objects. We take
H$_o$ = 73 km s$^{-1}$ Mpc$^{-1}$ for all luminosity comparisons. 

\begin{figure}
\plotone{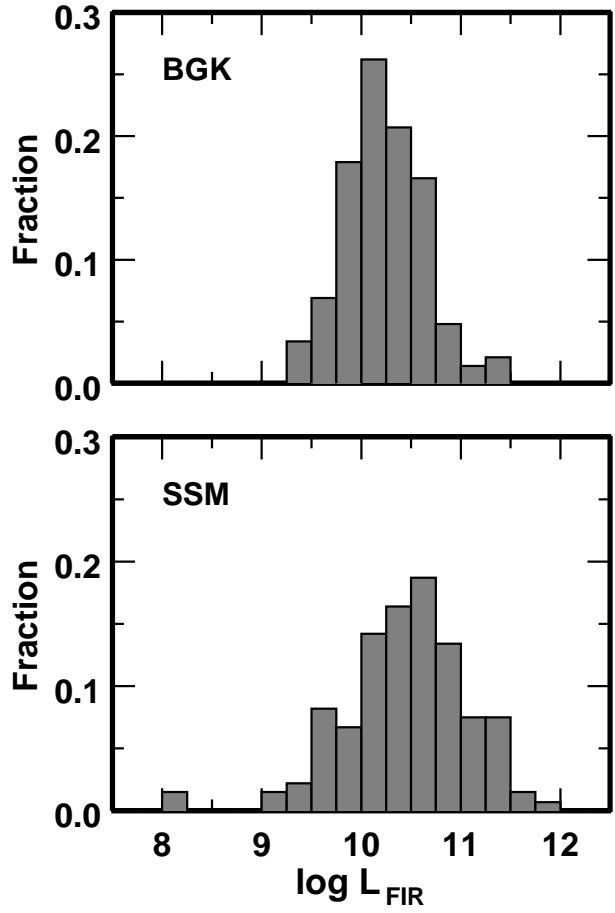}
\caption{Normalized distributions of L$_{FIR}$ for BGK and SSM pairs. Note
the extension of the SSM luminosity distribution into the ULIRG range.}
\label{fig:LFIR}
\end{figure}

Fig. \ref{fig:LFIR} shows that the range of log(L$_{FIR}$) in the SSM
sample is 9.1-11.8; in the BGK sample, the range is 9.3-11.4. 
As in Fig. \ref{fig:LH}, the core of the
SSM luminosity distribution is broader. 
The $\langle {\rm log} (L_{FIR}) \rangle$ in the SSM sample is
10.5 whereas in the BGK sample $\langle {\rm log} (L_{FIR}) \rangle$ is 10.2.
The shift
toward a broader distribution
with a greater $\langle {\rm log} (L_{FIR}) \rangle$ in the SSM
sample is as expected for an
IRAS-selected sample relative to a  B-selected sample. Neither sample contains ultraluminous infrared
galaxies.

\begin{figure}
\plotone{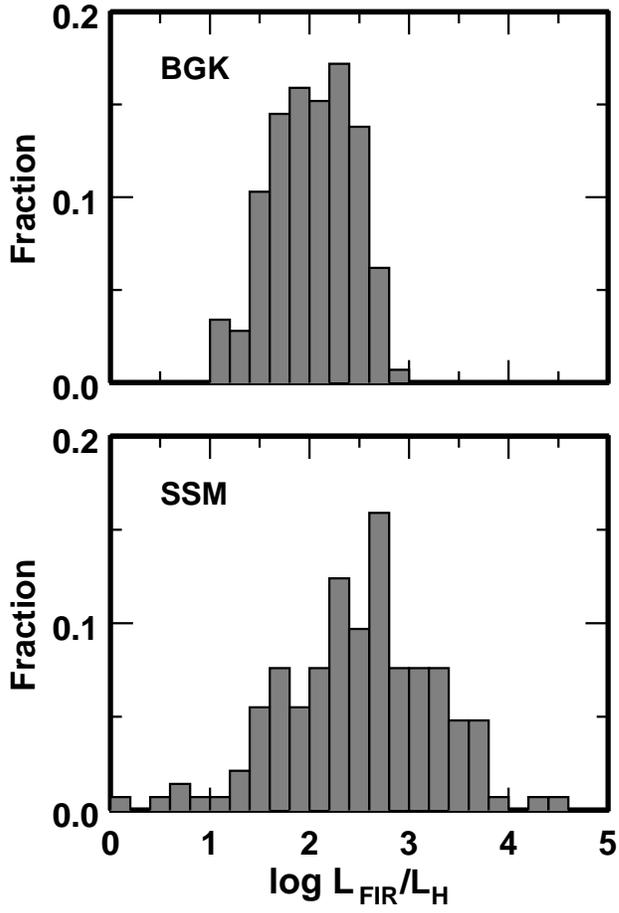}
\caption{Normalized distributions of L$_{FIR}$/L$_H$ for the BGK and SSM
pairs. }
\label{fig:LFIR/LH}
\end{figure}

Fig. \ref{fig:LFIR/LH} shows the distributions of ${\rm log\ L_{FIR}/L_H}$, a measure of the
normalized star formation rate, for both the SSM and BGK samples.
L$_{FIR}$ is directly proportional to the star formation
rate (Kennicutt 1988; Calzetti et al. 2000; Charlot et al. 2002; Kewley et al. 2002) and L$_H$ is roughly proportional to the stellar mass. Some
investigators use L$_K$ to normalize star formation rates. As we
have discussed, L$_K$ is affected by thermal emission from hot dust in these
systems and thus L$_H$ is a preferable proxy for the stellar mass.

In Fig. \ref{fig:LFIR/LH} , L$_H$ is the H-band luminosity for the pair derived from 2MASS
photometry.  Other investigators have shown that the IRAS
luminosities of close pairs are frequently  dominated by one of the
galaxies. In the BGK sample, nearly all of the pairs are unresolved in
IRAS and we have no way of knowing the detailed origin of the
IRAS emission. We thus normalize the IRAS luminosity by the total H-band
luminosity for the pair. This approach may systematically underestimate
the normalized star formation rate, but because most pairs are galaxies of
comparable H-band luminosity, the bias is of order a factor of 2.

The difference in the distributions is more pronounced in Fig.
\ref{fig:LFIR/LH} than in
Figure \ref{fig:LFIR}. The IRAS selection yields a sample rich
in pairs with ${\rm log\ L_{FIR}/L_H} \geq 2.8$. These objects are too faint at
B to be included in the BGK sample. The tail at    
${\rm log\ L_{FIR}/L_H} \leq 1$ in the SSM sample consist of low redshift
objects absent by construction from the BGK sample.

\begin{figure}
\plotone{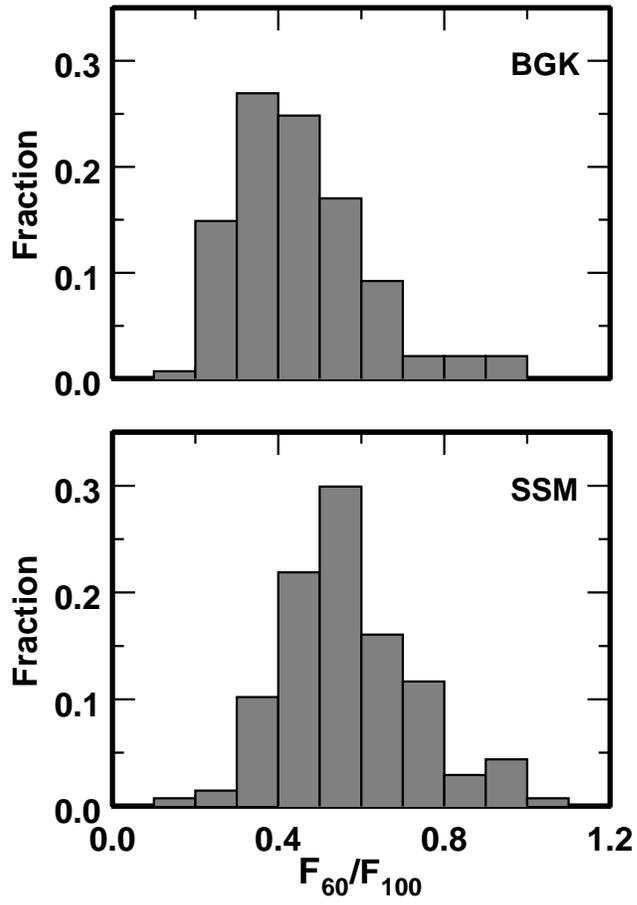}
\caption{Normalized distributions of F$_{60}$/F$_{100}$ for the BGK and SSM
pairs. }
\label{fig:F60/F100}
\end{figure}

The SSM IRAS selected pairs also contain somewhat hotter
dust than
the BGK pairs detected by IRAS. Figure \ref{fig:F60/F100} shows the distributions of
F$_{60}$/F$_{100}$ for both samples. The median temperature of the 
BGK pairs is 45K; the median for the SSM pairs is 50K.

In general, the BGK sample contains less extreme objects than the
SSM sample, but the overlap is substantial over two orders of
magnitude in both ${\rm L_{FIR}}$ and 
${\rm  L_{FIR}/L_H}$. Not surprisingly, pairs with
${\rm log\ L_{FIR}/L_H} \geq 2.8$ are rare
in a B-selected sample like BGK.  

\vfill \eject

\begin{deluxetable}{l c c c c c c c c}
\tablenum{1}
\tablecaption{2MASS Infrared Photometry}
\label{tbl-1}
\tablewidth{0pt}
\tablehead{
& & \multicolumn{3}{c}{7$^{\prime \prime}$ aperture} &
\multicolumn{3}{c}{Isophotal aperture} \\
\colhead{J2000 Designation} & \colhead{Redshift} & 
\colhead{J} & \colhead{H} & \colhead{K} &
\colhead{J} & \colhead{H} & \colhead{K} & 
\colhead{Split}}
\startdata
08042396+2930516 &  5447 & 13.87 & 13.22 & 12.90 & 13.51 & 12.91 & 12.60 & \\
08042496+2930236 &  5298 & 12.11 & 11.40 & 11.15 & 11.80 & 11.12 & 10.87 & \\
08065208+1844155 &  4557 & 12.03 & 11.32 & 11.05 & 11.72 & 11.02 & 10.74 & \\
08070665+1845506 &  4661 & 13.98 & 13.46 & 13.18 & 13.62 & 13.02 & 12.76 & \\
08100603+2455194 &  4128 & 12.70 & 11.97 & 11.71 & 12.19 & 11.47 & 11.24 & \\
08101117+2453344 &  4089 & 14.36 & 13.76 & 13.51 & 13.72 & 13.11 & 12.80 & \\
08111348+2512249 &  3995 & 12.82 & 12.10 & 11.80 & 11.35 & 10.66 & 10.40 & \\
08111591+2510459 &  4039 & 13.15 & 12.53 & 12.27 & 12.79 & 12.18 & 11.88 & \\
08112548+0853382 &  5763 & 14.09 & 13.45 & 13.15 & 14.31 & 13.66 & 13.34 & +\\
08112703+0856280 &  5704 & 14.00 & 13.32 & 13.05 & 13.86 & 13.19 & 12.97 & \\
\enddata
\tablenotetext{*}{Full table appears in the electronic edition}
\end{deluxetable}

\begin{deluxetable}{l c c c }
\tablenum{2}
\tablecaption{Journal of Observations}
\label{tbl-2}
\tablewidth{0pt}
\tablehead{
\colhead{UT Date}     & \colhead{Telescope} & 
\colhead{IR Imager}   & \colhead{Field of View}}
\startdata
13 Mar 2000--20 Mar 2000 & 1.2-m & STELIRCAM & 300\arcsec $\times$ 300\arcsec \\
23 Mar 2000--26 Mar 2000 & 2.1-m & ONIS  & 175\arcsec $\times$ 350\arcsec \\
15 Oct 2000--18 Oct 2000 & 2.1-m & SQIID & 300\arcsec $\times$ 300\arcsec \\
8 Apr 2001--12 Apr 2001 & 1.2-m & STELIRCAM & 300\arcsec $\times$ 300\arcsec \\
26 Sep 2001--27 Sep 2001 & 1.2-m & STELIRCAM & 300\arcsec $\times$ 300\arcsec \\
6 Oct 2001--9 Oct 2001 & 2.1-m & SQIID & 300\arcsec $\times$ 300\arcsec \\
3 Mar 2002--5 Mar 2002 & 1.2-m & STELIRCAM & 300\arcsec $\times$ 300\arcsec \\
21 Oct 2002--23 Mar 2002 & 1.2-m & STELIRCAM & 300\arcsec $\times$ 300\arcsec \\
\enddata
\end{deluxetable}
\clearpage

\begin{deluxetable}{l c c c c c c }
\tablenum{3}
\tablecaption{BGK Infrared Photometry}
\label{tbl-3}
\tablewidth{0pt}
\tablehead{
\colhead{J2000 Designation} & \colhead{$L$ (arcsec)} & 
\colhead{$w$ (arcsec)} & \colhead{PA (deg)} & \colhead{J} &
\colhead{H} & \colhead{K}}
\startdata
08042396+2930516 & 38.4 & 3.0 & 90.0 & 14.12 & 13.54 & 13.16 \\
08042496+2930236 & 4.0 & 3.0 & 90.0 & 13.28 & 12.60 & 12.34 \\
08065208+1844155 & 16.8 & 3.0 & 90.0 & 12.53 & 11.82 & 11.58 \\
08070665+1845506 & 28.3 & 3.0 & 90.0 & 14.90 & 14.28 & 14.03 \\
08111348+2512249 & 12.2 & 3.0 & 90.0 & 13.53 & 12.84 & 12.58 \\
08111591+2510459 & 28.8 & 3.0 & 90.0 & 13.87 & 13.22 & 13.03 \\
08112548+0853382 & 26.4 & 3.0 & 90.0 & 14.87 & 14.38 & 14.05 \\
08112703+0856280 & 28.8 & 3.0 & 90.0 & 14.32 & 13.62 & 13.41 \\
08184909+2113053 & 19.1 & 3.0 & 90.0 & 14.69 & 14.01 & 13.84 \\
08190189+2111093 & 4.3 & 3.0 & 90.0 & 15.31 & 14.58 & 14.35 \\
\enddata
\tablenotetext{*}{Full table appears in the electronic edition}
\end{deluxetable}

\begin{deluxetable}{l c r r r r }
\tablenum{4}
\tablecaption{IRAS Mid-Infrared Photometry}
\label{tbl-4}
\tablewidth{0pt}
\tablehead{
& & \multicolumn{4}{c}{Fluxes in Jy} \\
\colhead{J2000 Designation} & \colhead{IRAS Designation} & 
\colhead{12 $\mu$m} & \colhead{25 $\mu$m} & \colhead{60 $\mu$m} &
\colhead{100 $\mu$m}}
\startdata
08042396+2930516 & 08012+2939 & 0.25L & 0.37L & 0.43 &   1.10L \\
08042496+2930236 & 08012+2939 & 0.25L & 0.37L & 0.43 &   1.10L \\
08100603+2455194 & 08070+2503 & 0.48L & 0.31L & 0.93 &   1.13: \\
08101117+2453344 & 08070+2503 & 0.48L & 0.31L & 0.93 &   1.13: \\
08111348+2512249 & 08082+2521 & 0.25L & 0.46L & 2.14 &   5.97 \\
08111591+2510459 & 08082+2521 & 0.25L & 0.46L & 2.14 &   5.97 \\
08184909+2113053 & 08161+2120 & 0.25L & 0.61L & 0.54: & 1.51 \\
08190189+2111093 & 08161+2120 & 0.25L & 0.61L & 0.54: & 1.51 \\
08194129+2202311 & 08168+2211 & 0.25L & 0.34L & 0.74 &   2.12 \\
08194833+2201531 & 08168+2211 & 0.25L & 0.34L & 0.74 &   2.12 \\
\enddata 
\tablenotetext{*}{Full table appears in the electronic edition}
\tablenotetext{a}{To preserve associations of IRAS detections with
pair galaxies, we list IRAS associations for each galaxy in the pair.
This listing duplicates flux measurements for pair galaxies where 
IRAS could not spatially resolve the pair.  For the IRAS flxues, 
`L' indicates an upper limit (quality flag 1), and `:' indicates an 
approximate flux (quality flag 2).}
\end{deluxetable}
\clearpage

\begin{deluxetable}{l c c c c c}
\tablenum{5}
\tablecaption{BGK Pairs Brighter than Surace Flux Limits}
\label{tbl-5}
\tablewidth{0pt}
\tablehead{
\colhead{Pair ID} & \colhead{IRAS ID} &
\colhead{$F_{12}$ (Jy)} &
\colhead{$F_{25}$ (Jy)} &
\colhead{$F_{60}$ (Jy)} &
\colhead{$F_{100}$ (Jy)}}
\startdata
011934.70+032445.22 & 01171+0308 & 0.56: & 1.10 & 6.41 & 12.24 \\
023726.83+210716.52 & 02345+2053 & 0.56 & 1.22 & 10.18 & 16.93 \\
091549.24+405355.59 & 09120+4107 & 0.52 & 1.07 &  8.75 & 16.60 \\
160513.04+203230.75 & 16030+2040 & 0.24: & 0.87 & 7.04 & 10.10 \\
\enddata
\end{deluxetable}

\begin{deluxetable}{l c c c c c c c c}
\tablenum{6}
\tablecaption{JHK Model Parameters}
\label{tbl-6}
\tablewidth{0pt}
\tablehead{
\colhead{Model Number} & \colhead{(J-H)$_0$} & 
\colhead{(H-K)$_0$} & \colhead{$\sigma_{J-H}$} & 
\colhead{$\sigma_{H-K}$} & \colhead{$\sigma_B$} &
\colhead{$\sigma_D$} & \colhead{$\sigma_G$} &
\colhead{$\sigma_R$}}
\startdata
R & 0.67 & 0.27 & 0.05 & 0.05 & 0.00 & 0.000 & 0.00 & 0.020 \\
B & 0.67 & 0.27 & 0.05 & 0.05 & 0.05 & 0.000 & 0.00 & 0.020 \\
G & 0.67 & 0.27 & 0.05 & 0.05 & 0.00 & 0.000 & 0.05 & 0.020 \\
BGK  & 0.72 & 0.20 & 0.06 & 0.06 & 0.08 & 0.008 & 0.04 & 0.06 \\
\enddata
\end{deluxetable}
\clearpage

\end{document}